\pdfoutput=1
\IfFileExists{ptephy_v2.cls}
  {\documentclass[preprint,pteplogo]{ptephy_v2}\newif\ifptep\pteptrue}
  {\documentclass[11pt]{article}\usepackage[a4paper,margin=25mm]{geometry}\newif\ifptep\ptepfalse}
\ifptep
\preprintnumber{YITP-26-150}
\fi
\usepackage{amsmath,amssymb}
\usepackage{newtxtext}
\usepackage{newtxmath}
\usepackage{bm}
\usepackage{graphicx}
\usepackage{xcolor}

\newcommand{\nuc}[2]{\ensuremath{{}^{#1}\mathrm{#2}}}
\newcommand{\Cxii}{\nuc{12}{C}}
\newcommand{\Oxvi}{\nuc{16}{O}}
\newcommand{\Heiv}{\nuc{4}{He}}
\newcommand{\Caxl}{\nuc{40}{Ca}}

\newcommand{\zp}[1]{\ensuremath{0^{+}_{#1}}}

\newcommand{\threea}{\ensuremath{3\alpha}}

\newcommand{\amd}{AMD}
\newcommand{\gcm}{GCM}

\newcommand{\bg}{\ensuremath{\beta}--\ensuremath{\gamma}}

\newcommand{\Zi}[1]{\ensuremath{\bm{Z}_{#1}}}

\newcommand{\Bmat}{\ensuremath{B}}
\newcommand{\Wov}[1]{\ensuremath{\mathcal{W}_{#1}}}

\newcommand{\Rcen}[1]{\ensuremath{\bm{R}_{#1}}}
\newcommand{\kap}{\ensuremath{\kappa}}

\newcommand{\MeV}{\,\mathrm{MeV}}
\newcommand{\fm}{\,\mathrm{fm}}

\newcommand{\Volkov}{Volkov No.\,2}
\newcommand{\Fone}{F1}

\newcommand{\Fthreeb}{F3B$^{*}$}

\newcommand{\placefig}[2][\linewidth]{%
  \IfFileExists{figures/#2.pdf}{\includegraphics[width=#1]{figures/#2.pdf}}%
  {\fbox{\parbox[c][0.45\linewidth][c]{\dimexpr#1-2\fboxsep-2\fboxrule\relax}{\centering\ttfamily figures/#2.pdf\\[2pt]\small (not generated yet)}}}}

\newcommand{\src}[1]{\textcolor{blue!60!black}{\footnotesize[#1]}}
\newif\ifsrcnotes\srcnotesfalse
\ifsrcnotes\else\renewcommand{\src}[1]{\ignorespaces}\fi

\usepackage{booktabs}
\usepackage{braket}
\usepackage{mathrsfs}
\ifptep\else\usepackage[numbers,sort&compress]{natbib}\fi
\usepackage{hyperref}

\begin{document}

\title{Exact factorization of finite-range three-body matrix elements between Gaussian wave packets and its application to the $0^+$ states of \Cxii}

\author{Takatoshi Ichikawa}
\ifptep
\affil{Yukawa Institute for Theoretical Physics, Kyoto University, Kyoto 606-8502, Japan \email{ichikawa@yukawa.kyoto-u.ac.jp}}
\else
\date{}
\maketitle
\fi

\begin{abstract}
We introduce finite-range three-body interactions into the \bg-constrained generator coordinate method of antisymmetrized molecular dynamics. These genuine operators give saturation without an additional prescription for the projection, but require $\mathcal O(A^6)$ complex exponentials at every node of the projection grid. With the graph Laplacian of the ranges, as in spectral clustering widely used in machine learning, we factorize the exponential factor of a Gaussian three-body kernel exactly into pairwise tables of $\mathcal O(A^2)$ exponentials. A cofactor expansion rewrites the remaining contraction as complex matrix--matrix products. For \Cxii, the factorization reduces the time of the kernels with the three-body term from $530$ to $24$ times that of the two-body kernels. We apply the method to the $0^+$ states of \Cxii{} with the \Volkov, Tohsaki F1, and Kanada-En'yo--Akaishi F3B interactions, restricting the intrinsic states to time-reversal-invariant ones. Without the three-body terms of F1 and F3B, \Cxii{} collapses far below the \threea{} threshold, whereas with them the ground state lies close to its experimental position below the threshold. When we adjust the Majorana parameter to this position, the Hoyle state lies within $0.6\MeV$ of experiment for all three interactions.
\end{abstract}
\ifptep
\subjectindex{D10, D11, D12, D13}
\maketitle
\fi

\section{Introduction}
\label{sec:introduction}
Three-body interactions between nucleons have been introduced into microscopic models of nuclear structure because a two-body effective interaction alone does not give the saturation property. Self-consistent mean-field models based on energy density functionals \cite{Bender2003} supply the saturation by a density-dependent term $t_3\rho^\alpha$ with a fractional power $\alpha$. The two choices are not equivalent once the wave function is projected onto good angular momentum. The density-dependent term is not an operator and enters the mean-field equation through the rearrangement term. The angular-momentum projection and the generator coordinate method (\gcm) \cite{HillWheeler1953} require its matrix element between two different Slater determinants, which the functional does not define. The mixed-density prescription used in practice produces spurious divergences and finite steps in the energy projected onto good particle number \cite{Anguiano2001,Dobaczewski2007}, and the same problem arises in the configuration mixing \cite{Lacroix2009}. The projection of a density-dependent functional thus requires an additional prescription.

Microscopic cluster models instead use phenomenological three-body interactions of finite-range Gaussian form, whose matrix elements between Gaussian wave packets are analytic. Tohsaki proposed the F1 interaction, which supplements a three-range two-body interaction with a three-body term of the same ranges, and adjusted its parameters to the saturation property of nuclear matter and the $\alpha$-$\alpha$ scattering phase shifts \cite{Tohsaki1994}. Kanada-En'yo and Akaishi proposed the F3B interaction with a single-range three-body Gaussian of the three interparticle distances, and adjusted its Majorana parameter so that the antisymmetrized molecular dynamics (\amd) calculation combined with the \gcm{} reproduces the binding energies of the $N = Z$ nuclei from \Heiv{} to \Caxl{} \cite{KanadaEnyo2004F3B}. Because these interactions are genuine operators, their matrix elements between any two Slater determinants are defined, and the projection and the superposition need no additional prescription.

The \amd{} model can exploit this advantage. It optimizes a Slater determinant of Gaussian wave packets by energy variation \cite{Ono1992AMD,KanadaEnyo1995AMD,KanadaEnyo2012Review}, and because the rotated kernel at each node of the Euler-angle grid costs no more than the unprojected one, the energy variation after the parity and angular-momentum projections is practical \cite{KanadaEnyo1998VAP}. Suhara and Kanada-En'yo constrained the quadrupole deformation parameters $\beta$ and $\gamma$ of the intrinsic state and superposed the resulting Slater determinants by the \gcm{} \cite{Suhara2010}. This \bg-constrained \amd{} model described the Hoyle state, the second $0^+$ state of \Cxii, as a superposition of dilute \threea{} configurations spread over the \bg{} plane. We have also applied the replica-exchange Monte Carlo (RXMC) method \cite{Swendsen1986,Hukushima1996} to select the basis functions of the Brink $\alpha$-cluster model \cite{Brink1966} for the $0^+$ states of \Cxii{} \cite{Ichikawa2022RXMC}.
However, the calculations of Refs.~\cite{Suhara2010,Ichikawa2022RXMC} used the \Volkov{} interaction \cite{Volkov1965}, which does not have the saturation property. Such calculations prevent the collapse by fixing the width parameter of the packets and by adjusting the Majorana exchange parameter, or the strength of the spin-orbit interaction, to the binding energy of the nucleus of interest, and the same interaction then overbinds heavier nuclei \cite{Itagaki2016F1,Itagaki2016AQCM}. It remains unclear how the \bg-constrained \gcm{} describes the ground state and the \threea{} states of \Cxii{} with an interaction that has the saturation property, such as a three-body interaction.

The aim of this paper is to introduce three-body interactions into the \bg-constrained \gcm{} calculations, in which they have not yet been used because the direct evaluation of their matrix elements costs two or three orders of magnitude more than that of the two-body term. This cost has two bottlenecks. One is the number of complex exponentials. The three-body matrix element between two Slater determinants contains $A^3 \times A^3$ index tuples, each requiring an exponential of the wave-packet parameters, whereas the two-body term requires one exponential for each pair of packets. The \gcm{} calculation repeats this evaluation for $\mathcal O(10^4)$ pairs of basis functions and, for each pair, at every node of the Euler-angle grid, whose number exceeds $10^4$ for the dense grid required for dilute \threea{} configurations. The other is the form of the remaining arithmetic. Although the \amd{} matrix elements have simple forms, conventional \amd{} codes evaluate them in short loops over the packet indices, which do not form the large dense matrix--matrix products for which current processors are optimized. The throughput of the complex exponential therefore limits these codes to a small fraction of the peak performance.

To this end, we propose a method that reduces the cost of the three-body matrix elements between Slater determinants of Gaussian wave packets to about one order of magnitude above that of the two-body matrix elements. The method removes the two bottlenecks in two steps. The first step applies the graph Laplacian of spectral clustering, a method widely used in machine learning \cite{Luxburg2007}, to the three-body kernel. The exponent of a Gaussian three-body interaction is a sum of squared distances weighted by the ranges, i.e., a quadratic form with the graph Laplacian of the ranges. We show that the exponent of the kernel between two Slater determinants has the same form with new weights. The exponential factor of the kernel therefore factorizes exactly into pairwise factors, each of which depends on only two of the six single-particle indices. This reduces the number of exponential evaluations from $\mathcal O(A^6)$ to $\mathcal O(A^2)$. We then expand the three-body density, a $3\times3$ determinant, in cofactors along one of the ket indices and write the remaining contraction as a small number of complex matrix--matrix products. Their operation count is still $\mathcal O(A^6)$, but the optimized BLAS routine \textsc{zgemm} executes them at a much larger fraction of the peak performance than the direct evaluation.
A separable representation of a local finite-range $n$-body potential in a fixed one-particle basis is known as the exact tensor hypercontraction \cite{Parrish2013XTHC}. The present method instead factorizes the kernel between non-orthogonal Slater determinants of displaced Gaussian packets, whose centers change from one pair of determinants to the next and at every node of the angular-momentum projection.

We apply the method to \Cxii, where mean-field and cluster structures coexist. Its ground state contains a large component of the $p_{3/2}$ subshell-closed configuration, whereas the Hoyle state is a dilute \threea{} state near the \threea{} threshold \cite{Tohsaki2001THSR,Chernykh2007,KanadaEnyo2007}. We calculate the $0^+$ states with the \Volkov, Tohsaki F1, and Kanada-En'yo--Akaishi F3B interactions under a common width and spin-orbit interaction, restricting the intrinsic states to time-reversal-invariant ones. We adjust the Majorana parameter of each interaction to the ground state, as is customary with the Volkov interaction \cite{Itagaki2016AQCM}, and examine where the Hoyle state lies with respect to the \threea{} threshold.

The rest of this paper is organized as follows. In Sec.~\ref{sec:formalism}, we describe the \amd{} wave function and the three-body interactions. We then derive the factorization of the three-body matrix elements and describe the angular-momentum projection, the \gcm, and the intrinsic states obtained under a deformation constraint within the time-reversal-invariant space. In Sec.~\ref{sec:performance}, we present the measured computational cost. In Sec.~\ref{sec:c12}, we specify the three interactions, present the energy surfaces and the \gcm{} results for \Cxii, and examine the correlation excluded by the time-reversal restriction and the role of the three-body term. In Sec.~\ref{sec:summary}, we summarize this study and discuss the relation between the three-body interaction and the density dependence of the functionals. The Appendices give the proof that the inverse of the overlap matrix is the mean-field density matrix, the derivation of the three-body kernel, the gradient of the three-body term, and the two-body matrix elements.

\section{Formalism}
\label{sec:formalism}
\subsection{AMD wave function and matrix elements}
All matrix elements in this paper are taken between two Slater determinants, which differ in the projection and the \gcm. We here define the \amd{} wave function and write these matrix elements in terms of the transition density matrix. The \amd{} wave function is a Slater determinant of $A$ single-particle wave packets,
\begin{align}
  \ket{\Phi} = \frac{1}{\sqrt{A!}}\det\big[\psi_1,\ldots,\psi_A\big],
  \qquad
  \psi_i = \phi_{\Zi{i}}(\bm r)\,\chi_i\,\tau_i ,
  \label{eq:slater}
\end{align}
where $\phi_{\Zi{i}}$ is a Gaussian wave packet,
\begin{align}
  \phi_{\Zi{i}}(\bm r) = \Big(\frac{2\nu}{\pi}\Big)^{3/4}
  \exp\!\Big[-\nu\Big(\bm r - \frac{\Zi{i}}{\sqrt\nu}\Big)^2 + \frac{\Zi{i}^2}{2} - \frac{|\Zi{i}|^2}{2}\Big],
  \qquad
  \Zi{i} = \sqrt\nu\,\bm R_i + \frac{i\,\bm P_i}{2\hbar\sqrt\nu} .
  \label{eq:packet}
\end{align}
The complex vector $\Zi{i}$ represents the position $\bm R_i$ and the momentum $\bm P_i$ of the packet, and the width parameter $\nu$ is common to all packets. The spin wave function $\chi_i$ is a general spinor, $\chi_i = a_i\ket{\uparrow} + b_i\ket{\downarrow}$ with $|a_i|^2 + |b_i|^2 = 1$, and $\tau_i$ is the proton or neutron state. The variational parameters are $\{\Zi{i}, \chi_i\}$.

For a pair of Slater determinants $\Phi$ and $\Phi'$, the overlap matrix of the single-particle states is
\begin{align}
  \Bmat_{ij} = \braket{\psi_i|\psi'_j}
  = e^{\Wov{ij}}\,\braket{\chi_i|\chi'_j}\,\delta_{\tau_i\tau_j},
  \qquad
  \Wov{ij} = \overline{\Zi{i}}\cdot\bm Z'_j - \frac{|\Zi{i}|^2}{2} - \frac{|\bm Z'_j|^2}{2},
  \label{eq:bmatrix}
\end{align}
and the norm kernel is $\braket{\Phi|\Phi'} = \det\Bmat$. All matrix elements are expressed in terms of the one-body transition density matrix $\rho = \Bmat^{-1}$, which reduces to the Hartree--Fock density matrix of $\Phi$ for $\Phi' = \Phi$ (Appendix~\ref{app:density}). For one-, two-, and three-body operators,
\begin{align}
  \frac{\braket{\Phi|\hat V^{(1)}|\Phi'}}{\braket{\Phi|\Phi'}} &= \sum_{il} V_{i,l}\,\rho_{li},
  \label{eq:wick1}
  \\
  \frac{\braket{\Phi|\hat V^{(2)}|\Phi'}}{\braket{\Phi|\Phi'}} &= \frac{1}{2!}\sum_{ijlm} V_{ij,lm}\,
  \big(\rho_{li}\rho_{mj} - \rho_{mi}\rho_{lj}\big),
  \label{eq:wick2}
  \\
  \frac{\braket{\Phi|\hat V^{(3)}|\Phi'}}{\braket{\Phi|\Phi'}} &= \frac{1}{3!}\sum_{ijk}\sum_{lmn} V_{ijk,lmn}\,
  \mathcal D_{ijk,lmn},
  \qquad
  \mathcal D_{ijk,lmn} =
  \begin{vmatrix}
    \rho_{li} & \rho_{lj} & \rho_{lk} \\
    \rho_{mi} & \rho_{mj} & \rho_{mk} \\
    \rho_{ni} & \rho_{nj} & \rho_{nk}
  \end{vmatrix}.
  \label{eq:wick3}
\end{align}
Here $V_{i,l} = \braket{\psi_i|\hat v|\psi'_l}$, $V_{ij,lm} = \braket{\psi_i\psi_j|\hat v|\psi'_l\psi'_m}$, and $V_{ijk,lmn} = \braket{\psi_i\psi_j\psi_k|\hat v|\psi'_l\psi'_m\psi'_n}$ are the unsymmetrized kernels. The indices $i,j,k$ label the bra packets, and $l,m,n$ label the ket packets. The three-body density $\mathcal D_{ijk,lmn}$ is the $3\times 3$ determinant of the transition density matrix and contains all the direct and exchange terms. Equations~\eqref{eq:wick1}--\eqref{eq:wick3} hold for any pair of Slater determinants whose overlap matrix $\Bmat$ is nonsingular (Appendix~\ref{app:density}). In particular, they hold with $\Phi'$ replaced by $\hat R(\Omega)\Phi'$ in the angular-momentum projection described in Sec.~\ref{sec:projection}.

The Hamiltonian is
\begin{align}
  \hat H = \sum_i \hat t_i - \hat T_{\rm c.m.} + \sum_{i<j}\big(\hat v^{\rm cent}_{ij} + \hat v^{\rm LS}_{ij} + \hat v^{\rm Coul}_{ij}\big) + \sum_{i<j<k}\hat v^{(3)}_{ijk},
\end{align}
where $\hat T_{\rm c.m.}$ is the kinetic energy of the center of mass. The common width $\nu$ separates the center-of-mass motion, and we subtract $\hat T_{\rm c.m.}$ as its zero-point energy $3\hbar^2\nu/(2m)$, which is exact for states with vanishing total momentum. The central interaction is a sum of Gaussians with the Wigner, Bartlett, Heisenberg, and Majorana terms,
\begin{align}
  \hat v^{\rm cent}_{ij} = \sum_\alpha V^{(2)}_\alpha\,e^{-\kap_\alpha r_{ij}^2}\,
  \big(W_\alpha + B_\alpha \hat P^\sigma - H_\alpha \hat P^\tau - M_\alpha \hat P^\sigma\hat P^\tau\big),
  \label{eq:central}
\end{align}
where $\hat P^\sigma$ and $\hat P^\tau$ are the spin and isospin exchange operators, and the spin-orbit interaction is of the G3RS type \cite{Yamaguchi1979G3RS},
$\hat v^{\rm LS}_{ij} = \sum_{\alpha=1,2} u_\alpha\,e^{-\kap^{\rm LS}_\alpha r_{ij}^2}\,\hat P({}^3{\rm O})\,\bm L\cdot\bm S$, with the projector $\hat P({}^3{\rm O})$ onto the triplet-odd states. Appendix~\ref{app:twobody} gives the kernels of these interactions and of the Coulomb interaction.

\subsection{Three-body interactions}
\label{sec:interactions}
The three-body interactions considered in this paper are of two forms. The first is the three-body term of the Tohsaki interaction \cite{Tohsaki1994},
\begin{align}
  \hat v^{(3)}_{ijk}\big|_{\rm F1} = \frac13\sum_{\rm cyc}\sum_\alpha V^{(3)}_\alpha\,
  e^{-\kap_\alpha (r_{ij}^2 + r_{ik}^2)}\,
  \frac12\Big\{\big(W^{(3)}_\alpha + M^{(3)}_\alpha\hat P^r_{ij}\big)\big(W^{(3)}_\alpha + M^{(3)}_\alpha\hat P^r_{ik}\big) + (j \leftrightarrow k)\Big\},
  \label{eq:tohsaki}
\end{align}
where the cyclic sum runs over the choice of the apex nucleon $i$, $\hat P^r$ is the spatial exchange operator, and $W^{(3)}_\alpha = 1 - M^{(3)}_\alpha$. We call this the apex topology. The second is the three-body term of the F3B interaction \cite{KanadaEnyo2004F3B}. It is a single Gaussian in the distances of all three pairs without exchange operators,
\begin{align}
  \hat v^{(3)}_{ijk}\big|_{\rm F3B} = V^{(3)}\,e^{-\kap^{(3)}(r_{ij}^2 + r_{jk}^2 + r_{ki}^2)},
  \label{eq:kea}
\end{align}
which we call the triangle topology. We treat both interactions in a unified form in Sec.~\ref{sec:factorization}. The parameter values used in the application are given in Sec.~\ref{sec:parameters}.

\subsection{Factorization of the three-body matrix elements}
\label{sec:factorization}
The three-body term dominates the cost of the calculation. A direct evaluation of Eq.~\eqref{eq:wick3} requires one exponential for each of the $A^6$ index tuples, because the kernel of a Gaussian interaction is the exponential of a quadratic form in the six packet parameters. We reduce this cost in two steps. The first step uses the graph Laplacian of spectral clustering \cite{Luxburg2007}. The exponent of a Gaussian interaction is a quadratic form with this matrix, and the integration over the three nucleons gives the exponent of the kernel the same form with new weights. The quadratic form separates into terms each of which involves two packets, so that $\mathcal O(A^2)$ exponentials suffice. In the second step, we write the remaining sum over the tuples as matrix--matrix products. Appendix~\ref{app:kernel3} gives the derivations.

\paragraph{Kernel of a Gaussian three-body interaction}
Both Eqs.~\eqref{eq:tohsaki} and \eqref{eq:kea} are sums of terms of the form
\begin{align}
  \hat v = V\,\exp\!\Big[-\sum_{a<b}\kap_{ab}\,(\bm r_a - \bm r_b)^2\Big],
  \qquad a, b \in \{1, 2, 3\},
  \label{eq:general3}
\end{align}
with $\kap_{12} = \kap_{23} = \kap_{31} = \kap$ for the triangle topology and $\kap_{12} = \kap_{13} = \kap$, $\kap_{23} = 0$ for the apex topology. The exchange operators $\hat P^r$ in Eq.~\eqref{eq:tohsaki} permute the ket packets in the spatial factors and do not change this form (Appendix~\ref{app:kernel3}). The ranges enter the kernel through the $3 \times 3$ matrices
\begin{align}
  \mathsf K_{ab} = \delta_{ab}\sum_{c}\kap_{ac} - \kap_{ab},
  \qquad
  \mathsf A = 1 + \frac{\mathsf K}{2\nu},
  \qquad
  \mathsf G = \mathsf A^{-1},
  \qquad
  \widetilde V = V\,(\det\mathsf A)^{-3/2},
  \label{eq:kappa3}
\end{align}
with $\kap_{ba} = \kap_{ab}$ and $\kap_{aa} = 0$. The matrix $\mathsf K$ is the graph Laplacian of the ranges, $\sum_{a<b}\kap_{ab}(\bm r_a - \bm r_b)^2 = \sum_{ab}\mathsf K_{ab}\,\bm r_a\cdot\bm r_b$. Integrating over the coordinates of the three nucleons (Appendix~\ref{app:kernel3}), we obtain the kernel for the bra packets $(p_1, p_2, p_3) = (i, j, k)$ and the ket packets $(q_1, q_2, q_3) = (l, m, n)$,
\begin{align}
  V_{ijk,lmn} &= \widetilde V\,
  \exp\!\Big[-\sum_{a<b}2\mathsf G_{ab}\,\big(\Rcen{p_a q_a} - \Rcen{p_b q_b}\big)^2
  + \Wov{il} + \Wov{jm} + \Wov{kn}\Big]
  \notag\\
  &\quad\times
  \braket{\chi_i|\chi'_l}\braket{\chi_j|\chi'_m}\braket{\chi_k|\chi'_n}\,
  \delta_{\tau_i\tau_l}\delta_{\tau_j\tau_m}\delta_{\tau_k\tau_n}.
  \label{eq:kernel3}
\end{align}
Here $\Rcen{il} = (\overline{\Zi{i}} + \bm Z'_l)/2$ is $\sqrt\nu$ times the center of the transition density of the packets $i$ and $l$. The kernel is the product of the three overlaps $e^{\Wov{il}}$, multiplied by a Gaussian in the distances between the three transition centers. This Gaussian has the form of Eq.~\eqref{eq:general3} for the centers $\Rcen{p_a q_a}/\sqrt\nu$, with the ranges $\kap_{ab}$ replaced by the weights $2\nu\mathsf G_{ab}$. The weights reduce to $\kap_{ab}$ for $\kap \ll \nu$ and do not vanish for $\kap_{ab} = 0$ (Appendix~\ref{app:kernel3}). For vanishing ranges, $\mathsf G = 1$, and the kernel reduces to $V$ times the product of the overlaps.

\paragraph{Pairwise tables}
Writing the exponent of the Gaussian in Eq.~\eqref{eq:kernel3} with the graph Laplacian $1 - \mathsf G$ as $-2\sum_{ab}(\delta_{ab} - \mathsf G_{ab})\,\Rcen{p_a q_a}\cdot\Rcen{p_b q_b}$ and inserting $\Rcen{p_a q_a}$ and $\Wov{p_a q_a}$, we find (Appendix~\ref{app:kernel3}) that the exponent of the kernel consists of terms of three kinds,
\begin{align}
  \Xi_{p,q} &= \underbrace{\sum_{ab}\mathsf G_{ab}\,\overline{\bm Z}_{p_a}\!\cdot\bm Z'_{q_b}}_{\text{bra--ket}}
  \;\underbrace{-\,\frac12\sum_{ab}\big(\delta_{ab} - \mathsf G_{ab}\big)\,\overline{\bm Z}_{p_a}\!\cdot\overline{\bm Z}_{p_b}}_{\text{bra--bra}}
  \;\underbrace{-\,\frac12\sum_{ab}\big(\delta_{ab} - \mathsf G_{ab}\big)\,\bm Z'_{q_a}\!\cdot\bm Z'_{q_b}}_{\text{ket--ket}}
  \notag\\
  &\quad - \frac12\sum_a\big(|\bm Z_{p_a}|^2 + |\bm Z'_{q_a}|^2\big).
  \label{eq:epf}
\end{align}
Each term contains at most two of the six packets. The exponential of each term is thus an entry of an $A \times A$ table, for example $\exp(\mathsf G_{ab}\,\overline{\bm Z}_p\cdot\bm Z'_q)$ for all $p$ and $q$, and the kernel for any of the $A^6$ tuples is a product of such entries. We compute the tables once for each pair of Slater determinants, each range, and each pair of slots $(a, b)$. This requires $\mathcal O(A^2)$ exponentials instead of $\mathcal O(A^6)$, and the factorization is exact. To prevent overflow for extended configurations, we scale each table by a power of two, which introduces no rounding error.

\paragraph{Contraction by matrix--matrix products}
The tables remove the exponentials from the sum over the tuples but not its $A^6$ terms. We fix the bra triple $(i, j, k)$. By Eq.~\eqref{eq:epf}, the kernel is then a product of factors that depend on one ket index and of the ket--ket tables,
\begin{align}
  V_{ijk,lmn} &= C_{ijk}\,X_1(l)\,X_2(m)\,X_3(n)\,
  E_{12}(l, m)\,E_{13}(l, n)\,E_{23}(m, n),
  \notag\\
  E_{ab}(q, q') &= \exp\!\big[\mathsf G_{ab}\,\bm Z'_q\cdot\bm Z'_{q'}\big].
  \label{eq:fixed-bra}
\end{align}
Here $C_{ijk}$ collects $\widetilde V$, the bra--bra factors, and the normalization factors of the bra packets. The factor $X_a(q)$ collects the bra--ket factors, the diagonal ket--ket factor and the normalization factor of slot $a$, the spinor overlap, and the isospin factor, and it depends on the bra triple. We expand the three-body density along the ket index $n$, $\mathcal D_{ijk,lmn} = \sum_{c=1}^{3}\rho_{n p_c}\,\mathcal C_c(l, m)$, where $\mathcal C_c(l, m)$ is the signed $2\times 2$ minor of $\rho$ with the rows $l, m$ and the bra columns of the two slots other than $c$. The sum over $n$ in Eq.~\eqref{eq:wick3} then becomes the matrix product
\begin{align}
  Q_c(l, m) = \sum_n T_{(lm),\,n}\,Y_{n,\,c},
  \qquad
  T_{(lm),\,n} = E_{13}(l, n)\,E_{23}(m, n),
  \qquad
  Y_{n,\,c} = X_3(n)\,\rho_{n p_c},
  \label{eq:gemm3}
\end{align}
and the contribution of the bra triple is
\begin{align}
  \sum_{lmn} V_{ijk,lmn}\,\mathcal D_{ijk,lmn}
  = C_{ijk}\sum_{lm} X_1(l)\,X_2(m)\,E_{12}(l, m)\sum_{c}\mathcal C_c(l, m)\,Q_c(l, m).
  \label{eq:gemm3-sum}
\end{align}
The $A^2 \times A$ matrix $T$ contains only the ket--ket tables and does not depend on the bra triple. We stack the $A \times 3$ matrices $Y$ of a batch of bra triples column by column and compute $Q$ for the batch by one \textsc{zgemm} call for each range and exchange term. The isospin factors restrict each ket index to the isospin of the corresponding bra index. We therefore build $T$ and $Y$ within each of the eight isospin blocks $(\tau_1, \tau_2, \tau_3)$, and the inner dimension of the product is $A/2$ for $N = Z$. The product requires $24$ floating-point operations for each of the $A^6/8$ tuples allowed by the isospin factors, i.e., $\mathcal O(A^6)$ as before, and runs at the speed of the optimized library. The remaining sum in Eq.~\eqref{eq:gemm3-sum} requires $\mathcal O(A^5)$ operations in total.

In the angular-momentum projection (Sec.~\ref{sec:projection}), we rotate the ket packets and spinors by $\hat R(\Omega)$. We rebuild the tables and $\rho$ at each Euler node and perform the contraction once per node. The gradient of the projected energy, which the variation requires, consists of contractions of the same shape with the same matrix $T$ (Appendix~\ref{app:gradient}).

We have implemented the direct evaluation of Eq.~\eqref{eq:kernel3} for each tuple, the evaluation with the tables, and the matrix-product evaluation. The three agree to a relative difference smaller than $10^{-11}$ for all interactions and nuclei used in this paper. Section~\ref{sec:performance} uses the direct evaluation as the reference.

\subsection{Angular-momentum projection and GCM}
\label{sec:projection}
The intrinsic states break the rotational and reflection symmetries. We restore them by the angular-momentum and parity projections and superpose the projected states by the \gcm. We perform the projections numerically with the operator
\begin{align}
  \hat P^{J\pi}_{MK} = \frac{2J+1}{8\pi^2}\int d\Omega\, D^{J*}_{MK}(\Omega)\,\hat R(\Omega)\,\frac{1+\pi\hat\Pi}{2},
  \label{eq:projection}
\end{align}
where $\hat\Pi$ is the parity operator and $\hat R(\Omega)$ rotates the spatial coordinates and the spinors by the Euler angles $\Omega = (\alpha,\beta,\gamma)$. We use the product grid of the Fomenko rule for $\alpha$ and $\gamma$ \cite{Fomenko1970}, with equally spaced nodes and equal weights, and the Gauss--Legendre quadrature for $\beta$. Reference~\cite{Shimizu2023SO3} discusses these quadratures on SO(3) and their degrees of exactness. We also tested the Lebedev-type quadratures on SO(3) of Ref.~\cite{Shimizu2023SO3}, which require fewer nodes for the same degree of exactness. For strongly symmetry-broken \amd{} configurations, such as dilute \threea{} configurations, the expected gain was not obtained in practice. We evaluate the rotated kernels $\braket{\Phi|\hat H\hat R(\Omega)|\Phi'}$ and $\braket{\Phi|\hat R(\Omega)|\Phi'}$ at each node.

We superpose the intrinsic states $\Phi(\beta_a,\gamma_a)$ on the \bg{} plane, obtained in Secs.~\ref{sec:intrinsic} and \ref{sec:timereversal}, by the \gcm,
\begin{align}
  \ket{\Psi^{J\pi}_M} = \sum_{a}\sum_K c_{aK}\,\hat P^{J\pi}_{MK}\ket{\Phi(\beta_a,\gamma_a)},
\end{align}
and obtain the coefficients $c_{aK}$ and the energies by solving the Hill--Wheeler equation \cite{HillWheeler1953} with the Hamiltonian and norm kernels between the projected states. For the $0^+$ states studied here, $K = 0$ only.
Neighboring intrinsic states overlap strongly, and the norm kernel among them has small eigenvalues that are sensitive to the numerical error of the projection. We diagonalize the norm kernel $N_{ab} = \braket{\Phi(\beta_a,\gamma_a)|\hat P^{0+}_{00}|\Phi(\beta_b,\gamma_b)}$ normalized by its diagonal elements, $\tilde N_{ab} = N_{ab}/\sqrt{N_{aa}N_{bb}}$, and solve the Hill--Wheeler equation in the space of its eigenvectors with eigenvalues larger than the cutoff $10^{-4}$. We choose this cutoff above the numerical error of $\tilde N$ estimated for subsets of the states (Sec.~\ref{sec:setup}).

\subsection{Intrinsic states on the \texorpdfstring{\bg}{beta-gamma} plane}
\label{sec:intrinsic}
We generate the intrinsic states by energy variation after the $J^\pi = 0^+$ projection under a quadrupole-deformation constraint. We define $\beta$ and $\gamma$ from the mass quadrupole tensor of the intrinsic state. The definition is the same as in our previous work on the Brink $\alpha$-cluster model \cite{Ichikawa2022RXMC} and follows that used in the mean-field calculations \cite{Maruhn2014Sky3D},
\begin{align}
  Q_{kl} &= \operatorname{Re}\sum_{ij}\rho_{ji}\,\braket{\psi_i|\,3 x_k x_l - r^2\delta_{kl}\,|\psi_j}
  \notag\\
  &= \operatorname{Re}\,\frac{1}{\nu}\sum_{ij}\Bmat_{ij}\rho_{ji}\,\big(3\Rcen{ij,k}\Rcen{ij,l} - \Rcen{ij}^2\,\delta_{kl}\big),
  \qquad k, l \in \{x, y, z\}.
  \label{eq:quadrupole}
\end{align}
Here $\Rcen{ij}$ is that of Sec.~\ref{sec:factorization} with $\Phi' = \Phi$, measured from the center of mass, $\sum_i\operatorname{Re}\Zi{i} = 0$. The width term of the packets cancels between $3x_kx_l$ and $r^2\delta_{kl}$. The eigenvectors of $Q_{kl}$ define the principal axes of the intrinsic frame, and its eigenvalues are ordered as $Q_1 \le Q_2 \le Q_3$. The deformation parameters are
\begin{align}
  \beta\cos\gamma = \sqrt{\frac{\pi}{5}}\,\frac{Q_3}{A R_0^2},
  \qquad
  \beta\sin\gamma = \sqrt{\frac{\pi}{5}}\,\frac{Q_2 - Q_1}{\sqrt3\,A R_0^2},
  \qquad R_0 = 1.2\,A^{1/3}\fm ,
  \label{eq:betagamma}
\end{align}
i.e., $\beta\cos\gamma = \beta_{20}$ and $\beta\sin\gamma = \sqrt2\,\beta_{22}$, where $\beta_{2m} = 4\pi Q_{2m}/(5AR_0^2)$ are the dimensionless quadrupole moments in the principal-axis frame with $Q_{20} = \sqrt{5/16\pi}\,Q_3$ and $Q_{22} = \sqrt{5/96\pi}\,(Q_2 - Q_1)$. This definition differs from that of Ref.~\cite{Suhara2010}, which normalizes the quadrupole moments by the mean square radius of the state itself. With $R_0$ fixed by the mass number, $\beta$ measures deformation on an absolute length scale, and its values should not be compared directly with those of Ref.~\cite{Suhara2010}.
We impose the constraint similarly to Ref.~\cite{Ichikawa2022RXMC}, by minimizing
\begin{align}
  E' = E^{J=0} + \eta\Big[\Big(\frac{\beta_{20} - \beta_{20}^{t}}{s_{20}}\Big)^2 + \Big(\frac{\beta_{22} - \beta_{22}^{t}}{s_{22}}\Big)^2\Big] + \eta_0\,\nu|\bm R_G|^2 ,
  \label{eq:penalty}
\end{align}
where $E^{J=0}$ is the projected energy and $(\beta_{20}^t, \beta_{22}^t) = (\beta\cos\gamma, \beta\sin\gamma/\sqrt2)$ are the target values. The last term, with $\eta_0 = 10\MeV$ and the center of mass $\bm R_G = \sum_i\bm R_i/A$, removes the translation of the variational parameters. This translation leaves the energy and the deformation unchanged, because we evaluate both after shifting the packets to $\sum_i\operatorname{Re}\Zi{i} = 0$. We take $s_{20} = 1$ and $s_{22} = s_{20}/\sqrt2$. The penalty is then isotropic on the Cartesian shape plane $(\beta_{20}, \sqrt2\beta_{22})$, where the $(\beta,\gamma)$ grid of Sec.~\ref{sec:setup} is nearly uniformly spaced. We start from $\eta = 100\MeV$ and increase it by a factor of 10 until each term in the brackets of Eq.~\eqref{eq:penalty} becomes smaller than $10^{-3}$, i.e., $|\beta_{2m} - \beta_{2m}^t| < s_{2m}\sqrt{10^{-3}}$. Reference~\cite{Ichikawa2022RXMC} evaluated the gradient numerically. Here we use the analytic gradients of the projected energy and the penalty with respect to $\overline{\Zi{i}}$ and $\chi_i$. Appendix~\ref{app:gradient} gives the gradient of the projected three-body term. We minimize $E'$ by the limited-memory quasi-Newton method L-BFGS-B \cite{Byrd1995LBFGSB,Zhu1997LBFGSB} with loose bounds on the packet parameters.

\subsection{Time-reversal-invariant intrinsic states}
\label{sec:timereversal}
The variation of Sec.~\ref{sec:intrinsic} minimizes the energy of the $0^+$-projected state, whereas the constraint is imposed on the density of the intrinsic state. The projected energy $E^{J=0}$ is normalized by the norm of the $0^+$ component, $\braket{\Phi|\hat P^{0+}_{00}|\Phi}$, and does not depend on how small this component is. When the $0^+$ component is a small fraction of the norm, the intrinsic state can satisfy the constraint while its $0^+$ component can be close to the $0^+$ state of another deformation with a lower energy. This does not occur in the \bg-constrained \amd{} of Ref.~\cite{Suhara2010}, which the present method extends, or in the constrained \amd{} of Ref.~\cite{Kimura2007Mg31}. They vary the parity-projected state before the angular-momentum projection (VBP), and the state whose energy is minimized is the state whose density is constrained.

Without a further restriction, the present variation finds such states. On the fine grid of Sec.~\ref{sec:setup}, $\beta \le 0.6$, the intrinsic states break the spin saturation and acquire a net spin. When we suppress the net spin by a penalty, they have a large orbital current instead. The energy surface then shows spurious dips. At the deepest dips for \Volkov{} and \Fone, the energy is $2.5$--$4.2\MeV$ lower than at the neighboring grid points, the $0^+$ component is $7\times10^{-7}$ and $3\times10^{-8}$ of the norm, and its squared overlap with the $0^+$ state at the energy minimum is $0.92$ and $0.93$, respectively.  This mechanism does not depend on the interaction.

A net spin and a current are both odd under time reversal $\hat{\mathscr T}$. We therefore restrict the intrinsic states to those invariant under $\hat{\mathscr T}$, in which the spin and current densities vanish together. In each isospin block, the packets form pairs of a packet and its time-reversed image, $(\Zi{i}, \chi_i)$ and $(\overline{\Zi{i}}, -i\sigma_y\chi_i^{*})$, where $\sigma_y$ is the Pauli matrix. Because the two packets of a pair have the parameters $\Zi{i}$ and $\overline{\Zi{i}}$, $\sum_i\operatorname{Im}\Zi{i} = 0$, i.e., the total momentum vanishes, and the subtraction of $\hat T_{\rm c.m.}$ is exact. We vary one packet of each pair. Such pairs exist for even numbers of protons and neutrons, as in \Cxii. Every time-reversal-invariant Slater determinant of packets with the common $\nu$ has this paired form, because the time-reversed image of a packet is again a packet, packets with different centers are linearly independent, and $\hat{\mathscr T}^2 = -1$ for a nucleon.

\begin{figure}[tbp]
\centering
\placefig[\linewidth]{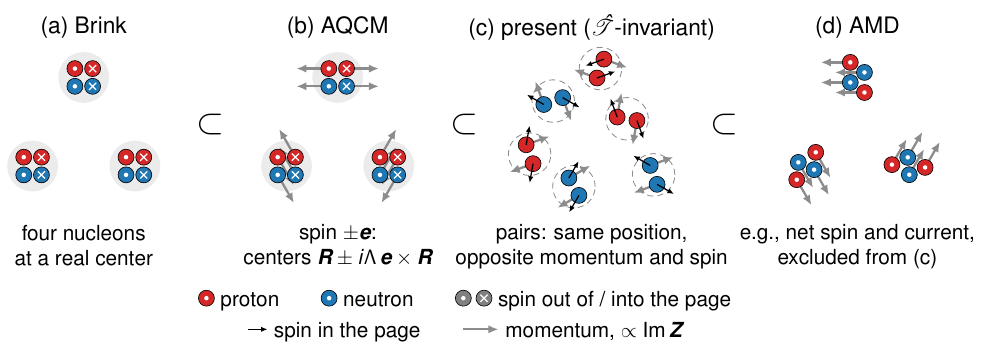}
\caption{Model spaces of the intrinsic states of \Cxii, nested as (a) $\subset$ (b) $\subset$ (c) $\subset$ (d). Each circle is a packet at its center $\operatorname{Re}\bm Z$. (a) Brink $\alpha$-cluster model. (b) AQCM: the nucleons with the spins $\pm\bm e$, with $\bm e$ pointing out of the page, move in opposite directions. $\bm R$ is the center of a quasi-$\alpha$ cluster, and $\Lambda$ is the dissociation parameter of the AQCM. (c) Present time-reversal-invariant space: the two packets of each pair have the same center and are drawn apart for visibility (dashed circles). (d) \amd: an example with a net spin and a current, which is excluded from (c). }
\label{fig:nesting}
\end{figure}

The present model space is the time-reversal-invariant part of the \amd{} space, in which this symmetry is usually not assumed \cite{Kimura2007Mg31}. It contains the Brink $\alpha$-cluster model \cite{Brink1966} and the antisymmetrized quasi-cluster model (AQCM) \cite{Itagaki2005AQCM,Suhara2013,Itagaki2016AQCM}, whose quasi-$\alpha$ clusters consist of such pairs, introduced as time-reversal partners in Ref.~\cite{MatsunoItagaki2017}. Figure~\ref{fig:nesting} shows these model spaces nested in each other. The present space does not group the pairs into clusters and varies their centers and spin directions freely.

With the restriction, the spurious dips disappear. The energies at the two grid points above rise by $4.4$ and $4.7\MeV$, and the $0^+$ components of all the restricted states on the fine grid are larger than $0.016$.  The restriction also excludes a correlation that does not arise from the constraint. At the minimum of the unrestricted energy surface, the intrinsic state has a small spin and is lower in energy than the restricted state, because the spin-up and spin-down packets of a pair separate in space (Sec.~\ref{sec:trexcluded}). Adding the time-reversed images to the \gcm{} basis would not recover this correlation, because the $0^+$ components of an unrestricted state and of its image overlap by more than $0.98$.  It is emphasized that the superpositions of the restricted states still span the whole space, as those of the unrestricted ones do, provided that the spin directions are varied freely. The antisymmetrized product of a packet and its time-reversed image, expanded in $\Zi{i}$, $\overline{\Zi{i}}$, the spinor amplitudes, and their complex conjugates, has coefficients that span all antisymmetric two-nucleon states, and the products of such pairs span all Slater determinants. The restriction thus slows the convergence of the \gcm{} rather than removing a part of the space. Section~\ref{sec:gcm} compares the present variation with the VBP for \Cxii{} with \Volkov.

\section{Performance}
\label{sec:performance}
\begin{table}[tb]
\centering
\caption{Cost of the three-body term of \Fone{} per evaluation of the kernel and its gradient for one pair of Slater determinants, i.e., one Euler node, on one thread for the test configurations of Sec.~\ref{sec:performance}, where the quantities are defined. The subscripts ref and fac denote the direct evaluation of Eq.~\eqref{eq:kernel3} and the factorized contraction of Eq.~\eqref{eq:gemm3}. $p$ is evaluated with respect to the nucleus in the preceding row. }
\label{tab:cost}
\small
\setlength{\tabcolsep}{3pt}
\begin{tabular}{lrrrrrrrrrr}
\toprule
 & $N_{\exp}^{\rm ref}$ & $N_{\exp}^{\rm fac}$ & $U_{\rm ref}$ & $U_{\rm fac}$ & $\tau_{\rm ref}$ & $\tau_{\rm fac}$ & $\tau_{\rm ref}/\tau_{\rm fac}$ & $U_{\rm fac}/\tau_{\rm fac}$ & $p_{\rm ref}$ & $p_{\rm fac}$ \\
 & & & (GFLOP) & (GFLOP) & (s) & (s) & & (GFLOP/s) & & \\
\midrule
\Cxii{}        & $2.2 \times 10^{6}$ & $7.9 \times 10^{3}$ & 1.76 & 0.19 & 0.177 & 0.0096 & $\times 18$ & 20 & --- & --- \\
\Oxvi{}        & $1.3 \times 10^{7}$ & $1.4 \times 10^{4}$ & 9.87 & 0.92 & 1.00  & 0.031  & $\times 33$ & 30 & 6.0 & 4.1 \\
\nuc{28}{Si}   & $3.6 \times 10^{8}$ & $4.3 \times 10^{4}$ & 284  & 21.8 & 29.0  & 0.44   & $\times 66$ & 49 & 6.0 & 4.8 \\
\Caxl{}        & $3.1 \times 10^{9}$ & $8.8 \times 10^{4}$ & 2410 & 172  & 260   & 2.80   & $\times 93$ & 62 & 6.2 & 5.2 \\
\bottomrule
\end{tabular}
\end{table}

\begin{figure}[tbp]
\centering
\placefig[0.55\linewidth]{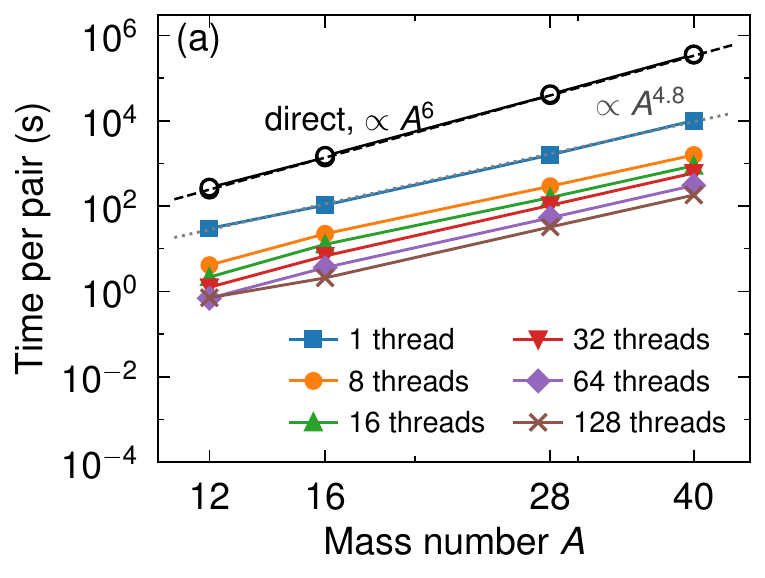}\hfill
\placefig[0.44\linewidth]{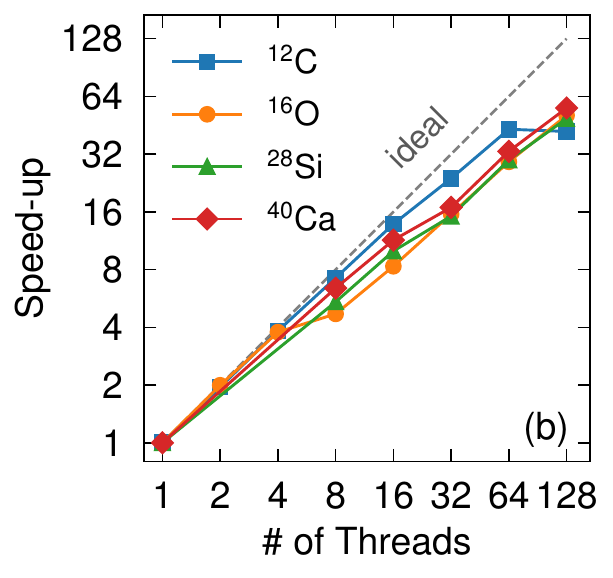}
\caption{(a) Time per pair of basis functions for the projected three-body kernel and its gradient of \Fone{} on the $1372$-node grid as a function of the mass number $A$, on $1$ and $8$ to $128$ threads. The open circles indicate the direct evaluation on one thread, estimated from Table~\ref{tab:cost} (see text). The dashed and dotted lines indicate $\propto A^{6}$ and $\propto A^{4.8}$. (b) Speed-up with respect to one thread. The dashed line indicates the ideal scaling. }
\label{fig:scaling}
\end{figure}

In this section, we examine how long the three-body kernels take with the factorization of Sec.~\ref{sec:factorization}, compared with the two-body kernels and with the direct evaluation, and how the gain changes with the mass number and the number of threads. We characterize the cost by the number of complex exponentials $N_{\exp}$, the number of useful floating-point operations $U$, the measured time $\tau$, and the exponent $p$ of $\tau \propto A^p$. $U$ does not include the exponentials. A complex multiplication counts $6$ operations and a complex fused multiply-add $8$, so that a \textsc{zgemm} product of an $m \times k$ and a $k \times n$ matrix counts $8mnk$. We perform the measurements on one compute node of the heian cluster at YITP, which has two AMD EPYC 9755 processors with $128$ cores each. The double-precision peak of one core is $131$ GFLOP/s at the maximum clock of $4.1$ GHz. We use GNU Fortran 14.2.1 and the BLAS of AOCL 5.2.0 on one thread. For \Cxii, \Oxvi, \nuc{28}{Si}, and \Caxl, Table~\ref{tab:cost} uses configurations of $3$, $4$, $7$, and $10$ nearly overlapping $\alpha$ clusters whose packet centers are displaced at random, and each time is the minimum of $20$ repetitions ($3$ for \Caxl).

For \Cxii, the factorization reduces the time of the kernels with the three-body term of \Fone{} from $530$ to $24$ times that of the two-body kernels. We measure the Hamiltonian and norm kernels of one pair of basis functions with the production code on one thread, on the grid of $34 \times 17 \times 34 = 19652$ nodes, which has half the nodes of the production grid of Sec.~\ref{sec:setup}. The kernels take $4.7\,$s with the two-body interaction alone (Appendix~\ref{app:twobody}), $2508\,$s with the direct three-body evaluation, and $115\,$s with the factorization. Table~\ref{tab:cost} includes the gradient, but these times do not. The times are proportional to the number of nodes within $6\%$ between the grids of $2916$ and $19652$ nodes, so that these ratios hold also on the production grid. With \Fthreeb, the kernels take $0.38$ times as long as with \Fone{} in the production calculations, because the triangle topology has a single range and a single term.

Table~\ref{tab:cost} shows that the two steps of Sec.~\ref{sec:factorization} remove the two bottlenecks of Sec.~\ref{sec:introduction} in different ways. In the direct evaluation, $N_{\exp}$ grows as $A^6$, whereas the factorized contraction evaluates exponentials only for the pairwise tables, $\propto A^2$. The ratio is $283$ for \Cxii{} and $3.5 \times 10^4$ for \Caxl. The tables also reduce $U$ by factors of $9$ to $14$, because the direct evaluation requires about $270$ operations per tuple and term for the exponent of Eq.~\eqref{eq:kernel3} and the $3 \times 3$ determinant, and the matrix product of Eq.~\eqref{eq:gemm3} requires only $24$. The direct evaluation runs at $9$--$10$ GFLOP/s, or $7$--$8\%$ of the peak, for all four nuclei, because it evaluates one exponential per tuple. The factorized contraction runs at $20$ and $62$ GFLOP/s, or $15$ and $47\%$ of the peak, for \Cxii{} and \Caxl, about one half of the library \textsc{zgemm} for matrices of the same shapes ($41$ and $119$ GFLOP/s). The throughput rises by factors of $2$ to $7$, and the two factors multiply to the speed-ups of $18$ for \Cxii{} and $93$ for \Caxl.

The speed-up grows with $A$, mainly because the throughput of the matrix product increases with its inner dimension $A/2$. Figure~\ref{fig:scaling}(a) shows the time per pair on the $1372$-node grid ($14 \times 7 \times 14$, sufficient for a spherical nucleus), measured with the OpenMP production code for the configurations of Table~\ref{tab:cost}. We see that the factorized time on one thread grows as $A^{4.8}$ and the direct time as $A^6$. For \Caxl, the factorized contraction takes $1.0 \times 10^4\,$s per pair, and the direct evaluation takes $3.6 \times 10^5\,$s. The direct times (open circles) are $\tau_{\rm ref}$ of Table~\ref{tab:cost} multiplied by the number of nodes. They are lower bounds, because the production code rebuilds the tables and rotates the packets at every node and takes $2.3$ to $2.6$ times as long per node as Table~\ref{tab:cost} for the factorized contraction. On $128$ of the $256$ cores, \Caxl{} takes $181\,$s per pair. A \gcm{} calculation for \Caxl{} with $10^3$ pairs thus costs $6 \times 10^3$ core hours with the factorization, whereas the direct evaluation would cost at least $10^5$ core hours even with ideal parallel scaling.

Figure~\ref{fig:scaling}(b) shows that the parallelization over the Euler nodes gives a further speed-up of $43$--$56$ on one compute node. For \Cxii, the speed-up is $90\%$ of the ideal value up to $8$ threads and saturates at $43$ on $64$ threads. This would be because $1372$ nodes give too little work per thread compared with the memory traffic of many threads. The larger workloads of \Oxvi, \nuc{28}{Si}, and \Caxl{} give $51$, $49$, and $56$ on $128$ threads.

Thus, the factorization reduces the time of the three-body kernels of \Cxii{} from $530$ to $24$ times that of the two-body kernels. Its speed-up grows from $18$ for \Cxii{} to $93$ for \Caxl, and the parallelization over the Euler nodes adds a factor of $43$--$56$ on one compute node, so that a \gcm{} calculation for \Caxl{} with $10^3$ pairs requires $6 \times 10^3$ core hours.

\section{Application to \texorpdfstring{\Cxii}{12C} with three interactions}
\label{sec:c12}
We apply the factorized three-body kernels to the \bg-constrained \gcm{} calculation of the $0^+$ states of \Cxii{} with three interactions, the \Volkov{} interaction without a three-body term and the Tohsaki F1 and F3B interactions with three-body terms. We ask where the Hoyle state lies once the ground state is fixed at its experimental position below the \threea{} threshold, and how the three-body term contributes to these energies. Section~\ref{sec:parameters} adjusts the Majorana parameter of each interaction to the ground state and fixes the other conditions in common, and Sec.~\ref{sec:setup} gives the grids and the basis. We then compare the energy surfaces in Sec.~\ref{sec:pes} and the $0^+$ states in Sec.~\ref{sec:gcm}. Section~\ref{sec:trexcluded} examines the correlation excluded by the time-reversal restriction, and Sec.~\ref{sec:threebody-role} the role of the three-body term.

\subsection{Interactions and parameters}
\label{sec:parameters}
We use the three interactions listed in Table~\ref{tab:force}. The first is the \Volkov{} interaction \cite{Volkov1965} with $B = H = 0.125$ as in Ref.~\cite{Suhara2010}. The second is the Tohsaki F1 interaction \cite{Tohsaki1994}, whose three-body term, Eq.~\eqref{eq:tohsaki}, has the same three ranges as its two-body term. The third is the F3B interaction \cite{KanadaEnyo2004F3B}, whose three-body term is Eq.~\eqref{eq:kea} and whose two-body term has three ranges, the third being a pure Wigner term. With the common spin-orbit interaction described below, we denote it by \Fthreeb.

For all three interactions, we adjust the Majorana parameter of the two-body term so that the \gcm{} calculation of Sec.~\ref{sec:gcm} reproduces the experimental energy of the ground state measured from the \threea{} threshold, $E(\zp{1}) - E_{\threea} = -7.275\MeV$, within $1\,$keV. Adjusting the Majorana parameter to the nucleus and to the model space is customary for effective interactions of this type. For example, Ref.~\cite{Itagaki2016AQCM} used $M = 0.62$ for \nuc{20}{Ne} and $0.65$ for \nuc{44}{Ti} with the Volkov interaction. Reference~\cite{KanadaEnyo2004F3B} calls the Majorana parameter of F3B ``an adjusting parameter in fine tuning to fit the experimental data'', and Ref.~\cite{Itagaki2016F1} reduced the three-body Majorana parameter of F1 from $1.909$ to $1.5$ to correct the overbinding of \Oxvi{} found in the $\alpha$-cluster model \cite{Itagaki1995}. Reference~\cite{Suhara2010} instead used $M = 0.60$ for \Volkov, which reproduces the $\alpha$-$\alpha$ scattering phase shifts, and adjusted the strength of the spin-orbit interaction to the ground state of \Cxii. Our adjustment changes $M$ by $+0.0132$ for \Volkov, $+0.0025$ for \Fone{} in all three ranges, and $-0.0069$ for \Fthreeb{} in the first two ranges (Table~\ref{tab:force}). The value for \Volkov{} is larger than $0.60$, because the ground state of Ref.~\cite{Suhara2010} lies below experiment when measured from the threshold and the present basis binds it further at the same $M$ (Sec.~\ref{sec:gcm}). We take the other parameters, including those of the three-body terms, from the original papers. We calculate the energy surfaces and the basis functions with the values of a preliminary adjustment, which differ from those of Table~\ref{tab:force} by at most $0.0005$.

To compare the three interactions on the same footing, we use a common width, a common spin-orbit interaction, and a common definition of the \threea{} threshold. We take $\nu = 0.235\fm^{-2}$ for all interactions. This is the value used in Ref.~\cite{Suhara2010} for \Cxii{} and is close to that of Ref.~\cite{Itagaki2016F1} for \Fone. For F3B, $E(\Heiv)$ at this width lies $0.72\MeV$ above its minimum, $-28.30\MeV$ at $\nu = 0.295\fm^{-2}$, which agrees with $28.2\MeV$ of Ref.~\cite{KanadaEnyo2004F3B} and checks our implementation of F3B. 

For the spin-orbit interaction, we take $u_1 = -u_2 = 1600\MeV$ and the ranges $\kap^{\rm LS} = 5.0$ and $2.778\fm^{-2}$ for all interactions as in Refs.~\cite{Suhara2010,SuharaPrivate,Itagaki2016F1}. The $p_{3/2}$ subshell closure of the \Cxii{} ground state is produced by the spin-orbit splitting, and a change of the spin-orbit interaction would dominate the differences in the central and three-body terms among the three interactions. Reference~\cite{KanadaEnyo2004F3B} used $u_1 = -u_2 = 1800\MeV$ with the ranges $5.0$ and $3.0\fm^{-2}$. We change the spin-orbit interaction only in the kernels and keep the basis. With the original strength at the present ranges, the adjustment of the Majorana parameter of F3B gives $M = 0.4213$, close to the original value $0.421$, and moves the \zp{2} state by $0.04\MeV$. For \Fone, with the same adjustment, the \zp{2} state moves by $0.04\MeV$ over $u_1 = 1400$--$1800\MeV$. The position of the \zp{2} state is thus nearly independent of the strength of the spin-orbit interaction. 

The \threea{} threshold is three times the energy of \Heiv{} calculated with the same $\nu$. The thresholds depend neither on the spin-orbit interaction, which does not contribute to the $(0s)^4$ configuration, nor on the Majorana parameter, because the space exchange is the identity for $(0s)^4$. The thresholds are $E_{\threea} = -82.760$, $-82.175$, and $-82.754\MeV$ for \Volkov, \Fone, and \Fthreeb, respectively, and we measure all energies of \Cxii{} in this section from them.

\begin{table}[tb]
\centering
\caption{Parameters of the interactions with the ranges $a_\alpha = \kap_\alpha^{-1/2}$. $M^{(2)}$ and $M^{(3)}$ are the Majorana parameters of the two- and three-body terms. $W_\alpha = 1 - M_\alpha$, and $B_\alpha = H_\alpha = 0$ unless listed. The Majorana parameters of the two-body terms are adjusted to the ground state of \Cxii{} (Sec.~\ref{sec:parameters}), and the original values (Ref.~\cite{Suhara2010} for \Volkov) are given in parentheses. For \Fthreeb, $1/\kap^{(3)} = 1.25\fm^2$ in Ref.~\cite{KanadaEnyo2004F3B}. }
\label{tab:force}
\small
\begin{tabular}{lccc}
\toprule
 & range 1 & range 2 & range 3 \\
\midrule
\multicolumn{4}{l}{\Volkov{} (two-body only, $B = H = 0.125$)} \\
\quad $a$ (fm)          & 1.80     & 1.01    & --- \\
\quad $V^{(2)}$ (MeV)   & $-60.65$ & $61.14$ & --- \\
\quad $M^{(2)}$         & 0.6132 (0.60) & 0.6132 (0.60) & --- \\
\midrule
\multicolumn{4}{l}{\Fone} \\
\quad $a$ (fm)          & 2.5     & 1.8      & 0.7 \\
\quad $V^{(2)}$ (MeV)   & $-5.00$ & $-43.51$ & $60.38$ \\
\quad $M^{(2)}$         & 0.7525 (0.75) & 0.4645 (0.462) & 0.5245 (0.522) \\
\quad $V^{(3)}$ (MeV)   & $-0.31$ & $7.73$   & $219.0$ \\
\quad $M^{(3)}$         & 0       & 0        & 1.909 \\
\midrule
\multicolumn{4}{l}{\Fthreeb} \\
\quad $a$ (fm)          & 1.2       & 0.7      & 0.9 \\
\quad $V^{(2)}$ (MeV)   & $-198.34$ & $300.86$ & $22.5$ \\
\quad $M^{(2)}$         & 0.4141 (0.421) & 0.4141 (0.421) & 0 \\
\quad $B$               & 0.10      & 0.10     & 0 \\
\quad $H$               & 0.085     & 0.085    & 0 \\
\quad $a^{(3)}$ (fm)    & 1.118     & ---      & --- \\
\quad $V^{(3)}$ (MeV)   & 600       & ---      & --- \\
\bottomrule
\end{tabular}
\end{table}

\subsection{Numerical setup}
\label{sec:setup}
Unless otherwise noted, the energy surfaces of Sec.~\ref{sec:pes} and the \gcm{} of Secs.~\ref{sec:gcm} and \ref{sec:threebody-role} use one set of intrinsic states for each interaction. We obtain the states by the variation of Secs.~\ref{sec:intrinsic} and \ref{sec:timereversal} on two grids of the \bg{} plane. The fine grid covers $\beta \le 0.6$ with $\Delta\beta = 0.05$, and its $\gamma$ points are equally spaced along the arc, with a step of $5^\circ$ at $\beta = 0.6$. The coarse grid covers $\beta \le 5$ with $\Delta\beta = 0.25$ and a step of $2.5^\circ$ at $\beta = 5$. The two grids have 91 and 273 points. The projections, in the variation and in the \gcm{} kernels, use the grid of $34 \times 34 \times 34$ nodes. By increasing the number of nodes, we have confirmed for subsets of the states that the energies are converged.  We start the variation at each grid point from two initial states and adopt the lower of the two results.

The \gcm{} basis of each interaction consists of the states of the fine grid and those of the coarse grid with $0.6 < \beta \le 3.75$, $91 + 154 = 245$ states, and the cutoff of Sec.~\ref{sec:projection} leaves $184$--$195$ dimensions. We show the states of the coarse grid with $\beta \ge 4$ in the energy surfaces but do not use them in the \gcm. With more nodes for subsets of the states, we find that the quadrature error of $\tilde N$ is at most $4.7\times10^{-5}$ for $\beta \le 3.75$, below the cutoff, but reaches $1.6\times10^{-4}$ when the states with $\beta = 4$ are included. 

The low-lying states are nearly converged with respect to the range of $\beta$: extending the basis from $\beta \le 3.5$ to $\beta \le 3.75$ changes $E(\zp{1}) - E_{\threea}$ by less than $1\,$keV and $E(\zp{2}) - E_{\threea}$ by at most $0.02\MeV$. It changes the ground-state radius by less than $0.001\fm$, the \zp{2} radius by at most $0.03\fm$, and the monopole matrix element of Sec.~\ref{sec:gcm} by at most $0.07\fm^2$. The \zp{3} state is not converged, and the extension lowers its energy by $0.07$--$0.17\MeV$ and increases its radius by $0.07$--$0.14\fm$. Lowering the cutoff to $10^{-6}$ lowers the ground state by $0.03$--$0.12\MeV$ and changes the excitation energy of the \zp{2} state by at most $0.05\MeV$. 

\subsection{Energy surfaces on the \texorpdfstring{\bg}{beta-gamma} plane}
\label{sec:pes}
\begin{figure}[tbp]
\centering
\placefig[0.48\linewidth]{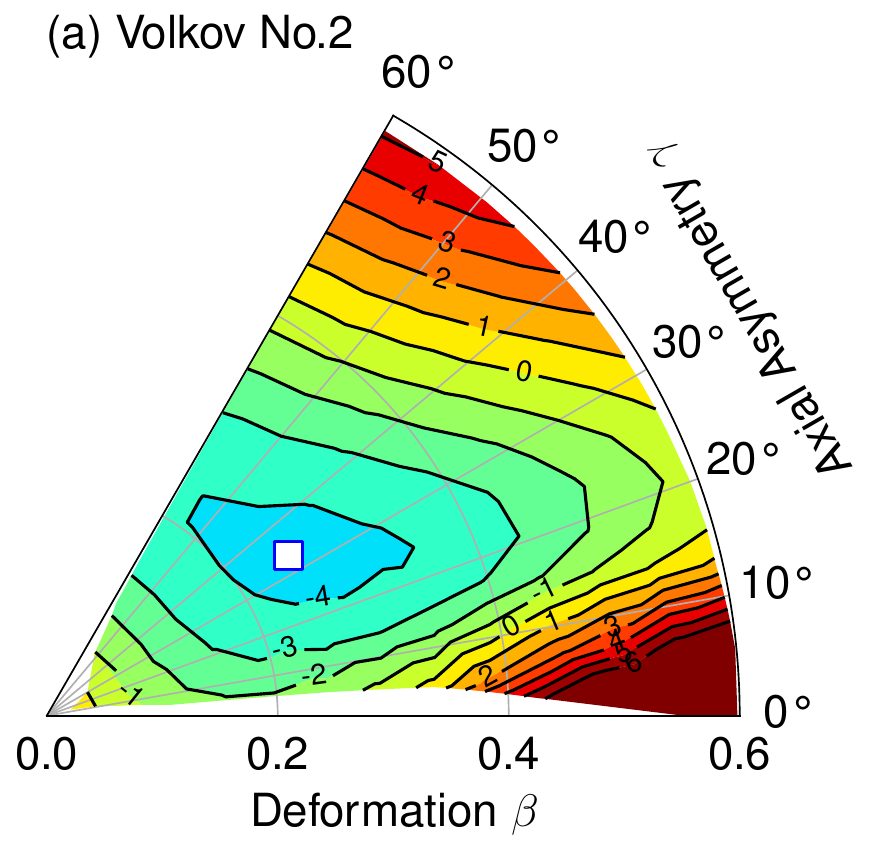}\hfill
\placefig[0.48\linewidth]{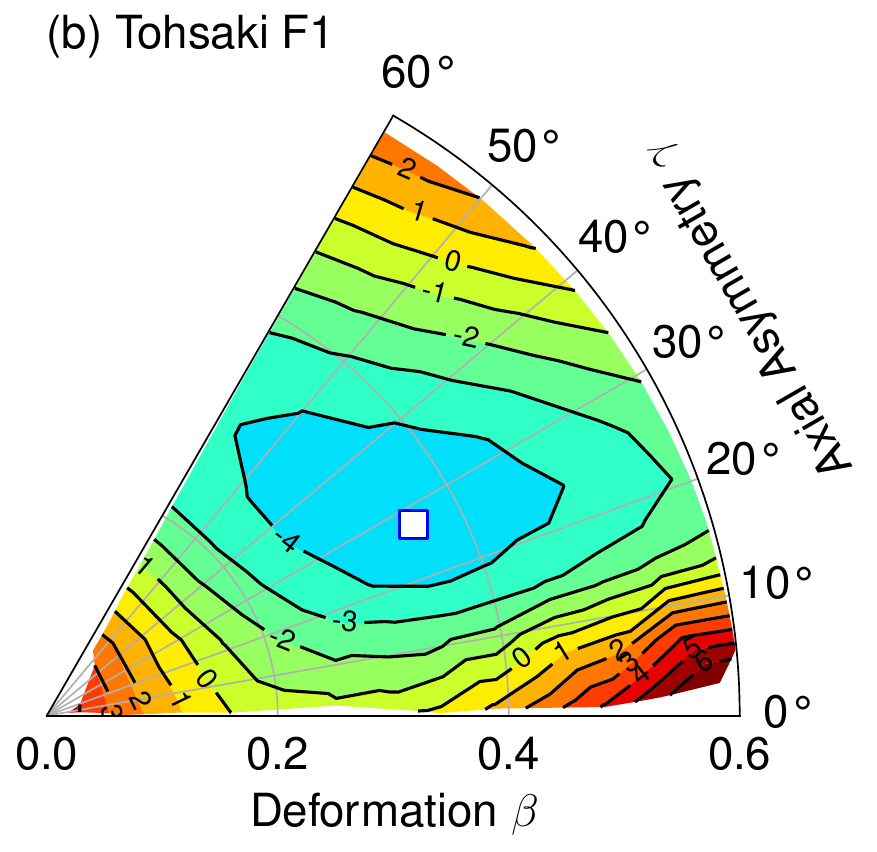}\\[4pt]
\placefig[0.48\linewidth]{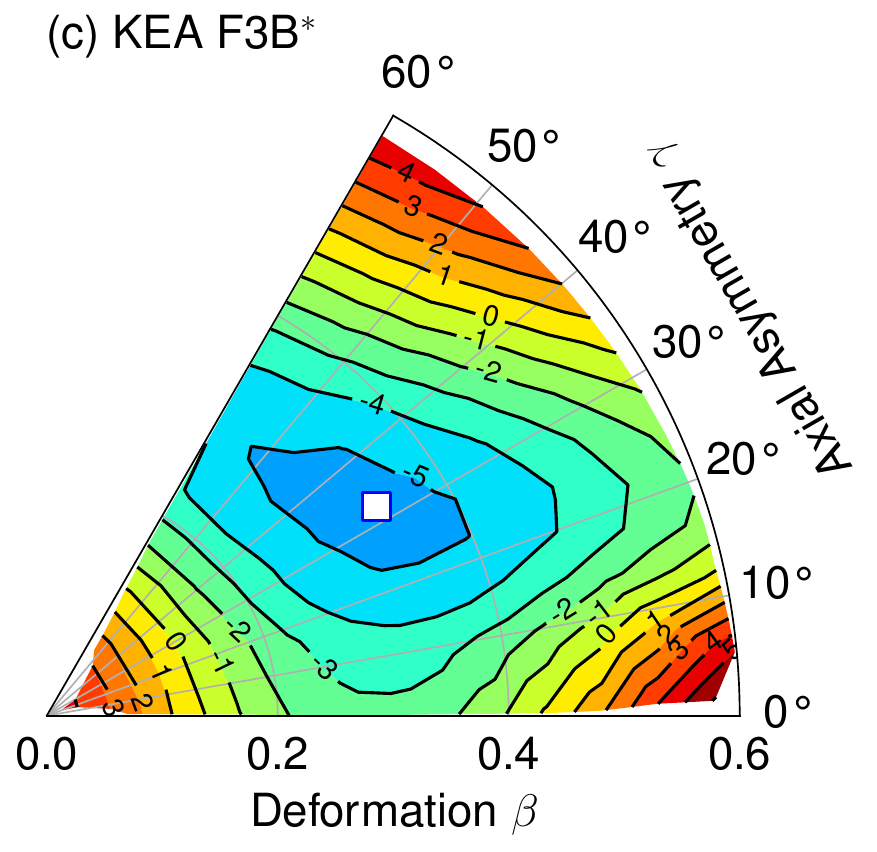}
\caption{Energy surfaces on the \bg{} plane for the $0^+$ state of \Cxii, measured from the \threea{} threshold, for the (a) \Volkov, (b) \Fone, and (c) \Fthreeb{} interactions on the fine grid (Sec.~\ref{sec:setup}). The surfaces are drawn at the values of $\beta$ and $\gamma$ of the varied states. The contour lines are drawn every $1\MeV$. The color scale is saturated above $+6\MeV$, where no contour lines are drawn. The open squares indicate the minima. }
\label{fig:pes}
\end{figure}
\begin{figure}[tbp]
\centering
\placefig[0.48\linewidth]{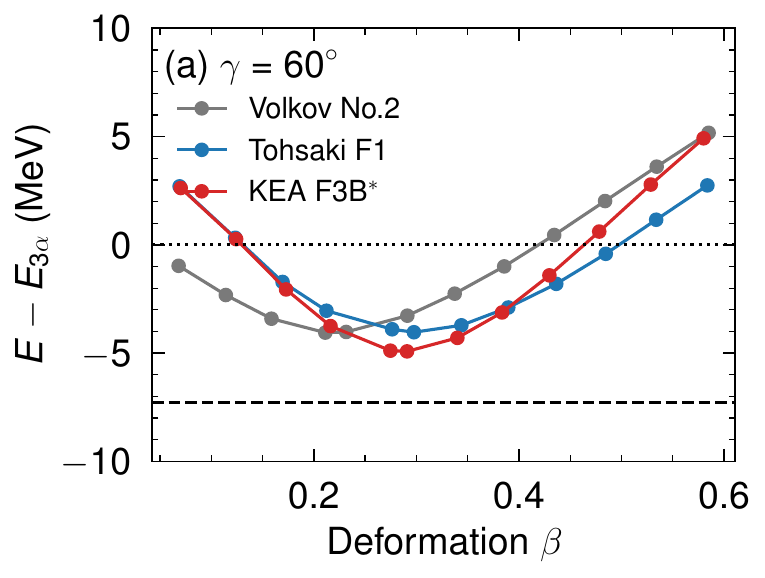}\hfill
\placefig[0.48\linewidth]{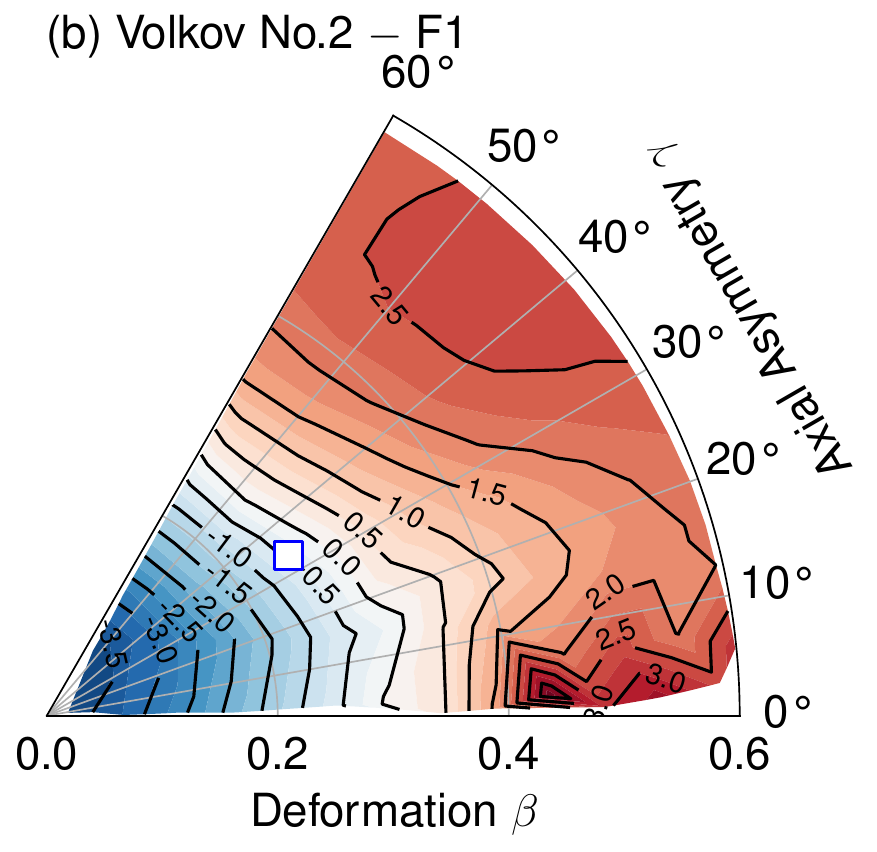}\\[4pt]
\placefig[0.48\linewidth]{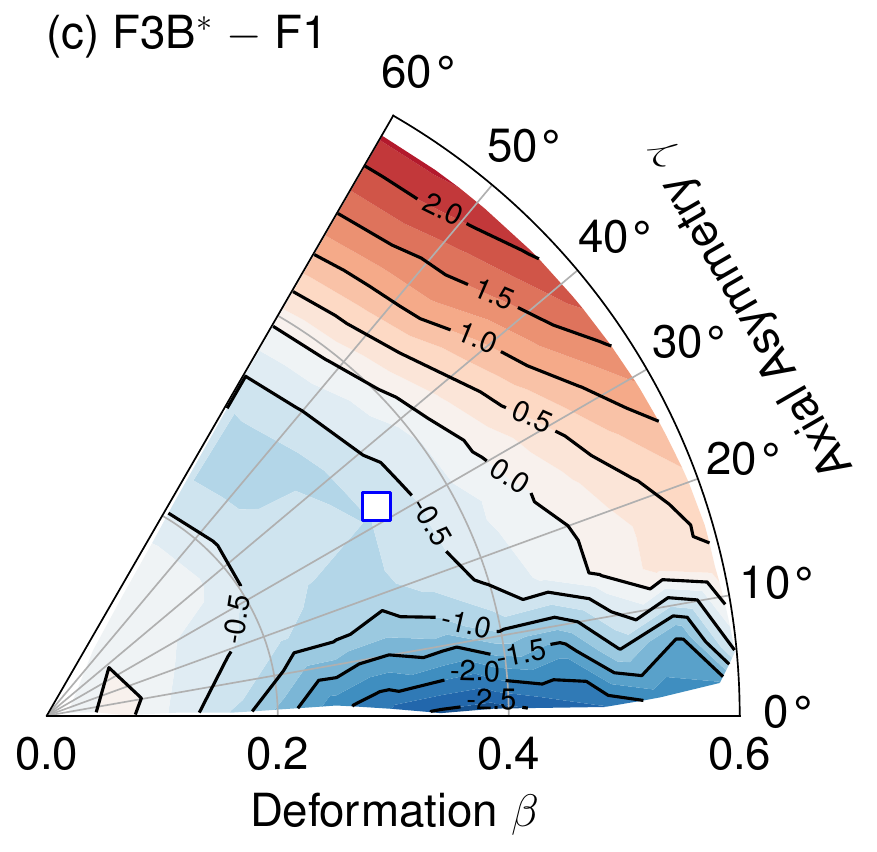}
\caption{(a) Energies of the states of Fig.~\ref{fig:pes} constrained to $\gamma = 60^\circ$ as functions of their $\beta$. The dashed and dotted lines indicate the experimental ground-state energy and the threshold. (b), (c) Differences of the surfaces of \Volkov{} and \Fthreeb{} from that of \Fone, evaluated at the $\beta$ and $\gamma$ of the \Fone{} states, or of the other interaction where the \Fone{} state lies outside its surface, by linear interpolation. The contour lines in (b) and (c) are drawn every $0.5\MeV$. The open squares indicate the minima of the \Volkov{} and \Fthreeb{} surfaces. }
\label{fig:pescut}
\end{figure}
\begin{figure}[tbp]
\centering
\placefig[0.48\linewidth]{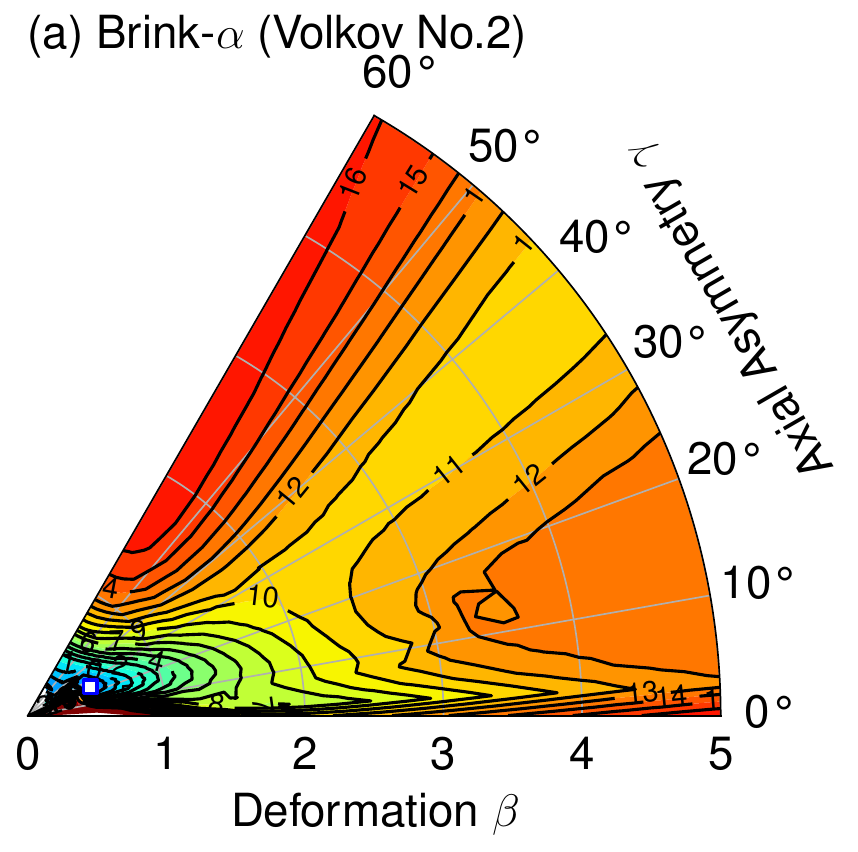}\hfill
\placefig[0.48\linewidth]{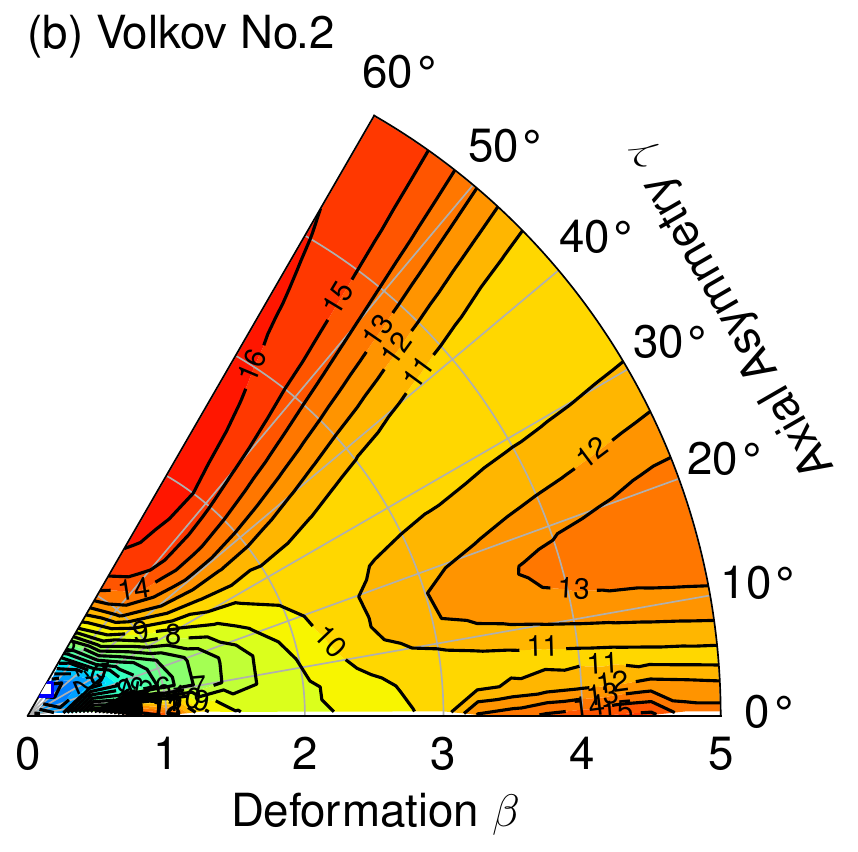}\\[4pt]
\placefig[0.48\linewidth]{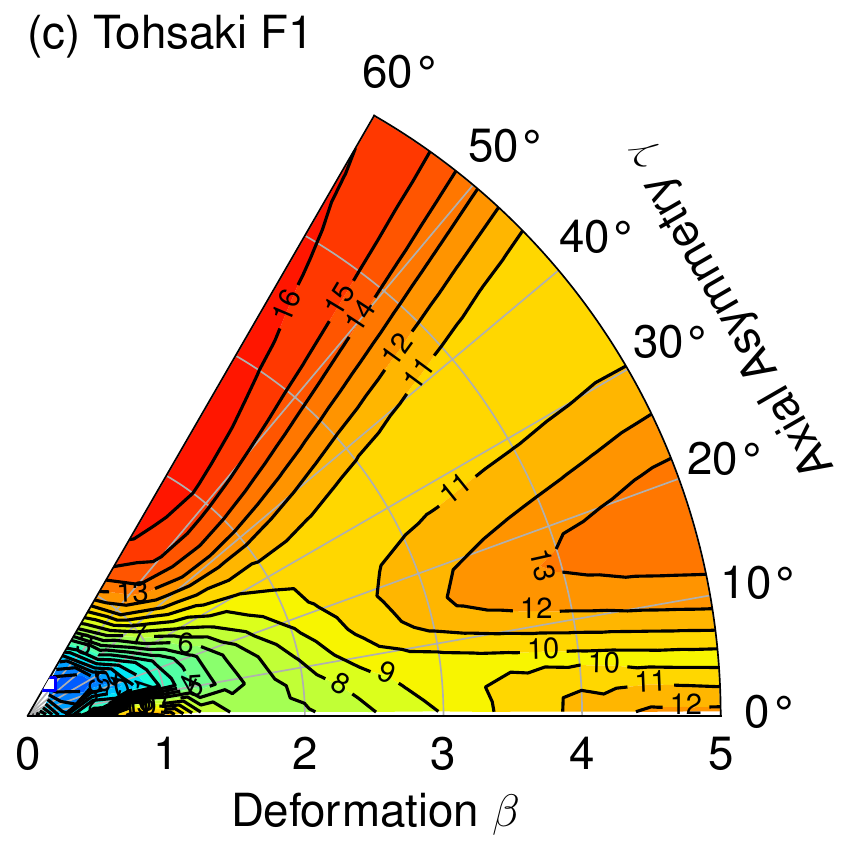}\hfill
\placefig[0.48\linewidth]{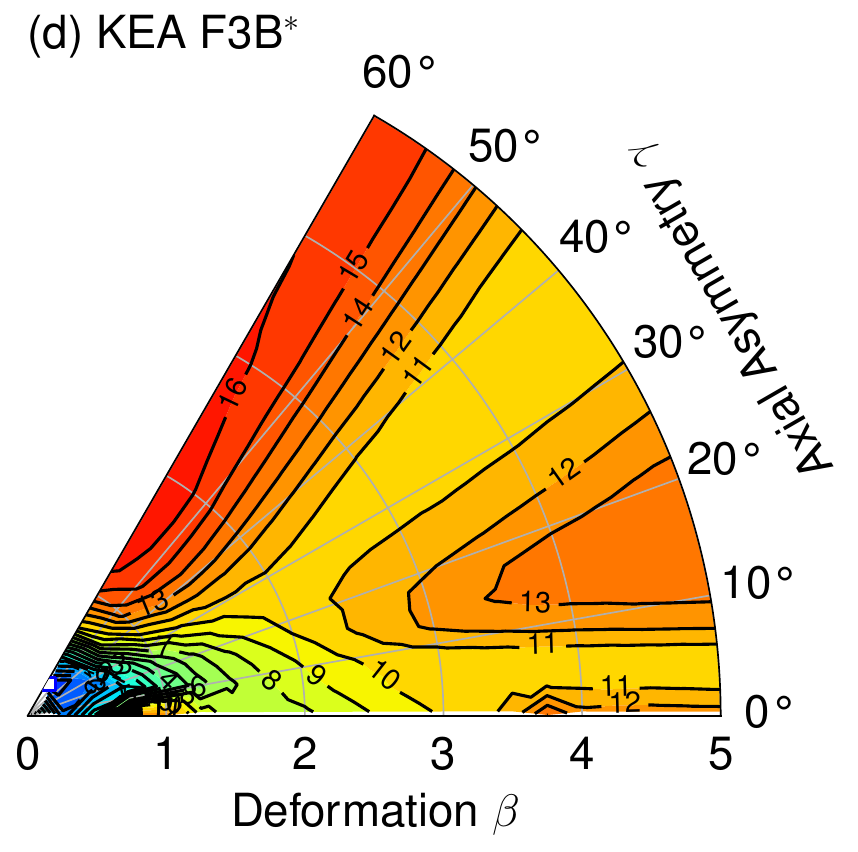}
\caption{(a) Energy surface of the Brink $\alpha$-cluster model with the \Volkov{} interaction ($M = 0.60$) \cite{Ichikawa2022RXMC}, measured from its \threea{} threshold. (b)--(d) The same as Fig.~\ref{fig:pes} but for $\beta \le 5$ on the coarse grid. In all panels, the contour lines are drawn every $1\MeV$, and the color scale is saturated above $+20\MeV$. }
\label{fig:peswide}
\end{figure}
We compare in this subsection how much the energy surfaces of the three interactions differ around the ground state on the fine grid of Sec.~\ref{sec:setup} and in the wide region on the coarse grid.

On the fine grid (Fig.~\ref{fig:pes}), the minima of the three surfaces lie $1.8$--$2.9\MeV$ above the experimental ground-state energy, $-7.275\MeV$. They are at $(\beta, \gamma) = (0.25, 34^\circ)$, $(0.36, 28^\circ)$, and $(0.34, 32^\circ)$, at $-4.39$, $-4.76$, and $-5.43\MeV$ for \Volkov, \Fone, and \Fthreeb, respectively. No state on the $\gamma = 60^\circ$ cut reaches the experimental energy either [Fig.~\ref{fig:pescut}(a)]. The \gcm{} of Sec.~\ref{sec:gcm} recovers this difference as the correlation energy.

We see that the three surfaces have the same valley structure. The \Volkov{} surface is the shallowest. It lies $0.97\MeV$ above \Fone{} at the minimum of \Fone. It lies below \Fone{} for $\beta \lesssim 0.25$, by up to $3.7\MeV$ at $\beta = 0$, and above it for larger $\beta$, by up to $3.9\MeV$ [Fig.~\ref{fig:pescut}(b)]. The \Fthreeb{} surface lies $0.58\MeV$ below \Fone{} at the minimum of \Fone{} and differs from it by $-2.6$ to $+2.3\MeV$ over the surface [Fig.~\ref{fig:pescut}(c)]. This difference varies smoothly with deformation: \Fthreeb{} is higher near $\gamma = 60^\circ$ at $\beta \gtrsim 0.45$ and lower along $\gamma = 0^\circ$.

Figure~\ref{fig:peswide} shows that the three \amd{} surfaces have the same landscape also in the wide region up to $\beta = 5$. The energy rises steeply from the minimum to $\beta \simeq 1.5$ and then becomes flat along $\gamma = 60^\circ$. Up to $\beta = 5$, it remains between $15.8$ and $16.7\MeV$ above the \threea{} threshold for all three interactions. Near $\gamma = 0^\circ$ ($\gamma \le 12^\circ$), in contrast, the lowest energy rises up to $\beta \simeq 3.25$ and then stays at $9.5$--$10.7\MeV$, about $6\MeV$ below the flat region along $\gamma = 60^\circ$. The surface of the Brink $\alpha$-cluster model with the same definition of $\beta$ and $\gamma$ [Fig.~\ref{fig:peswide}(a)] \cite{Ichikawa2022RXMC} has a flat region at the same height, $+16.1$ to $+16.7\MeV$. It differs from the \amd{} surfaces near the ground state, where its minimum is shallow because the $\alpha$ clusters are frozen, and near $\gamma = 0^\circ$, where it lies up to $3.6\MeV$ above the \Volkov{} surface.

The height of the flat region is mostly the zero-point kinetic energy of the relative motion of the three $\alpha$ clusters. At large $\beta$, the configurations along $\gamma = 60^\circ$ consist of three localized $\alpha$ clusters at the corners of an equilateral triangle, whose side is $6.5\fm$ at $\beta = 1.5$ and $12\fm$ at $\beta = 5$. Each relative coordinate contributes $\hbar^2\nu/(2m) = 4.9\MeV$ per dimension. The $J = 0$ projection removes the three rotational degrees of freedom, but the three vibrational ones remain, $3\hbar^2\nu/(2m) = 14.6\MeV$. The Coulomb energy of the three point charges is $2.6\MeV$ at $\beta = 1.5$ and $1.4\MeV$ at $\beta = 5$. Adding it gives the estimate $+16.1$ to $+17.3\MeV$, which reproduces the height and its slow decrease within $1.2\MeV$. This flat region would correspond to the \threea{} continuum described by localized wave packets. The \gcm{} superposition over $\beta$ and $\gamma$ recovers the correlation energy of this relative motion and gives the \zp{2} state near the threshold (Sec.~\ref{sec:gcm}).

We conclude that the three interactions give surfaces of the same shape, which differ from \Fone{} by up to $4\MeV$ around the ground state and by less than $1\MeV$ in the flat region of the \threea{} configurations.

\subsection{GCM results}
\label{sec:gcm}
\begin{table}[tb]
\centering
\caption{Energies of the $0^+$ states of \Cxii{} and the \threea{} threshold (MeV), and the number $N_{\rm b}$ of basis functions of the \gcm. The middle part uses the original Majorana parameters with the same basis. The experimental values are taken from Refs.~\cite{AME2020,Kelley2017A12}. The lower part lists the VBP \bg-\amd{} \cite{Suhara2010} and the Brink-$\alpha$ RXMC \cite{Ichikawa2022RXMC} calculations with the \Volkov{} interaction ($M = 0.60$). The \zp{2} energy of the VBP calculation is read from Fig.~11 of Ref.~\cite{Suhara2010}, and its threshold is calculated in the present work with the same interaction and $\nu$. }
\label{tab:gcm}
\small
\setlength{\tabcolsep}{3.5pt}
\begin{tabular}{lccccccc}
\toprule
 & $E(\zp{1})$ & $E(\zp{2})$ & $E_{\threea}$ & $E(\zp{1}) - E_{\threea}$ & $E(\zp{2}) - E_{\threea}$ & $E_x(\zp{2})$ & $N_{\rm b}$ \\
\midrule
\multicolumn{8}{l}{Majorana parameter adjusted} \\
\Volkov{}            & $-90.034$ & $-81.791$ & $-82.760$ & $-7.274$ & $+0.969$ & 8.243 & 245 \\
\Fone{}              & $-89.449$ & $-81.952$ & $-82.175$ & $-7.274$ & $+0.223$ & 7.497 & 245 \\
\Fthreeb{}           & $-90.028$ & $-81.796$ & $-82.754$ & $-7.274$ & $+0.958$ & 8.232 & 245 \\
\midrule
\multicolumn{8}{l}{Original Majorana parameters} \\
\Fone{}              & $-90.057$ & $-82.228$ & $-82.175$ & $-7.882$ & $-0.053$ & 7.828 & 245 \\
\Fthreeb{}           & $-88.837$ & $-81.279$ & $-82.754$ & $-6.084$ & $+1.475$ & 7.559 & 245 \\
\midrule
Exp.               & $-92.162$ & $-84.508$ & $-84.887$ & $-7.275$ & $+0.379$ & 7.654 & --- \\
\midrule
\bg-\amd{} (VBP) \cite{Suhara2010}          & $-92.5$  & $-82.61$ & $-82.760$ & $-9.74$ & $+0.15$ & 9.89 & 196 \\
Brink-$\alpha$ RXMC \cite{Ichikawa2022RXMC} & $-87.96$ & $-81.37$ & $-82.71$  & $-5.25$ & $+1.34$ & 6.59 & 100 \\
\bottomrule
\end{tabular}
\end{table}

\begin{figure}[tbp]
\centering
\placefig[0.7\linewidth]{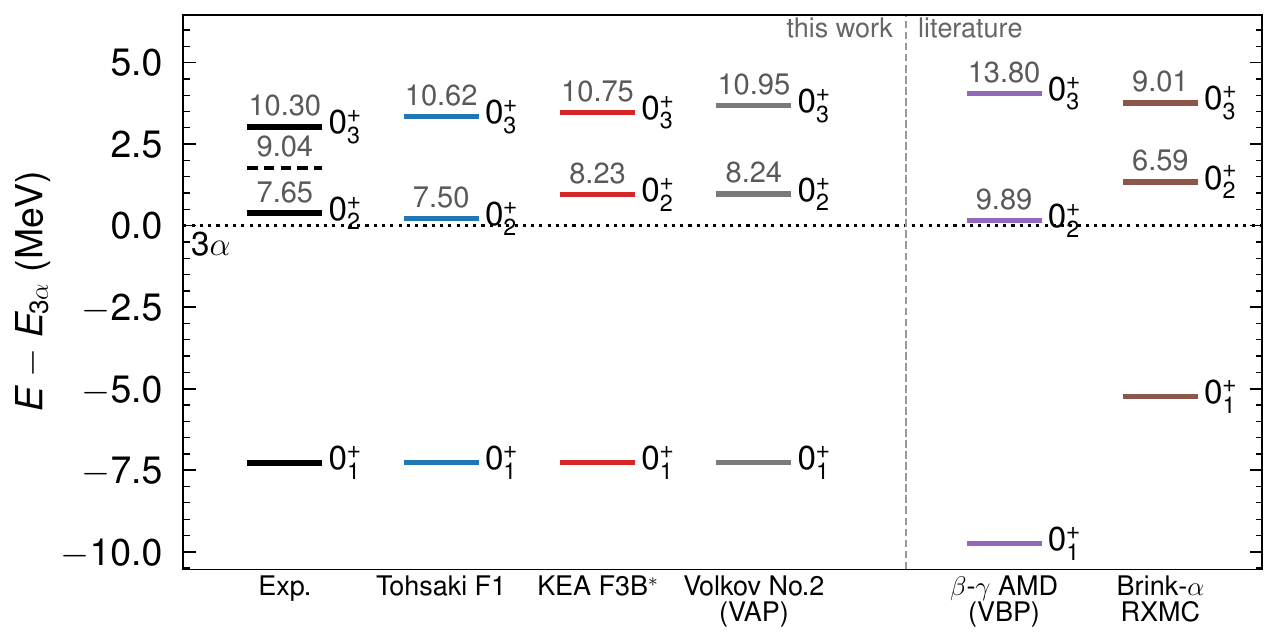}
\caption{Energy levels of the $0^+$ states measured from the \threea{} threshold (dotted line) with the Majorana parameters adjusted to the experimental position of the ground state, compared with experiment. Right of the dashed line are the calculations in the lower part of Table~\ref{tab:gcm}. VAP denotes the present variation after the projection. The numbers indicate the excitation energies in MeV. The dashed bar at $9.04\MeV$ indicates the lower of the two $0^+$ components into which Ref.~\cite{Itoh2011} decomposes the broad $0^+$ state. }
\label{fig:levels}
\end{figure}
We now ask where the \zp{2} state lies and how large it is for each of the three interactions. We also compare the results with those of the original parameters and of other calculations with the \Volkov{} interaction.

With the adjusted Majorana parameters, the \zp{2} state lies $0.97$, $0.22$, and $0.96\MeV$ above the threshold for \Volkov, \Fone, and \Fthreeb, respectively, against $0.38\MeV$ in experiment (Table~\ref{tab:gcm} and Fig.~\ref{fig:levels}). The excitation energies are $8.24$, $7.50$, and $8.23\MeV$, against $7.654\MeV$ \cite{Kelley2017A12}. With \Volkov{} and \Fthreeb, the Hoyle state lies at the same position, $0.58$--$0.59\MeV$ above experiment, whereas with \Fone{} it lies $0.16\MeV$ below. 

With the original Majorana parameters, the excitation energy of \Fthreeb, $7.56\MeV$, is closer to experiment, but this agreement follows from the error of the ground state, which is underbound by $1.19\MeV$. For \Fone, the ground state is overbound by $0.61\MeV$, and the excitation energy is $7.83\MeV$. The Majorana parameter shifts the energy of the ground state $2.1$--$2.3$ times as much as that of the \zp{2} state: $dE/dM = 177$ and $85\MeV$ for \Fthreeb, $238$ and $107\MeV$ for \Fone, and $291$ and $125\MeV$ for \Volkov{} in the \zp{1} and \zp{2} states. The excitation energy is therefore smaller when the ground state is underbound. Fixing the ground state increases the excitation energy of \Fthreeb{} by $0.67\MeV$ and decreases that of \Fone{} by $0.33\MeV$. We compare the \zp{2} states at the fixed ground state. 

The right part of Fig.~\ref{fig:levels} and the lower part of Table~\ref{tab:gcm} compare the result of \Volkov{} with two calculations using the same \Volkov{} interaction with $M = 0.60$. The Brink $\alpha$-cluster model with basis functions selected by the RXMC method \cite{Ichikawa2022RXMC} gives the \zp{2} state at $+1.34\MeV$, $0.37\MeV$ above the present $+0.97\MeV$. Its ground state is underbound by $2\MeV$ because the $p_{3/2}$ subshell closure is absent. The AQCM \cite{Suhara2013} restores this closure within the time-reversal-invariant space (Sec.~\ref{sec:timereversal}).

With the same Hamiltonian, the present variation after the projection gives lower energies than the VBP of Ref.~\cite{Suhara2010} (Sec.~\ref{sec:timereversal}), more for the compact ground state than for the dilute \zp{2} state. The \bg-\gcm{} calculation of Ref.~\cite{Suhara2010} gives a ground state $2.5\MeV$ deeper and a \zp{2} state $0.8\MeV$ lower than the present calculation, and an excitation energy of $9.89\MeV$. The two calculations differ in both the Majorana parameter ($0.60$ versus $0.6132$) and the basis. Their contributions can be separated because the Hamiltonian kernel is linear in $M$ and the norm kernel does not depend on it. With a basis generated at $M = 0.60$ on the same grids, the kernels at $M = 0.6132$ give $E(\zp{1}) - E_{\threea} = -7.38$ and $E(\zp{2}) - E_{\threea} = +0.90\MeV$, and those at $M = 0.60$ give $-11.54$ and $-1.26\MeV$. Changing $M$ by $0.013$ shifts the ground state by $4.2\MeV$ and the \zp{2} state by $2.2\MeV$ with this basis. At the same $M = 0.60$, this basis is more bound than the VBP basis of Ref.~\cite{Suhara2010} by $1.8\MeV$ for the ground state and by $1.4\MeV$ for the \zp{2} state, and its excitation energy is larger, $10.28$ compared with $9.89\MeV$. The \gcm{} over the \bg{} plane thus does not remove the difference between the two kinds of intrinsic states. 

\begin{table}[tb]
\centering
\caption{Point-nucleon rms radii (fm) and the monopole transition matrix element $M(E0; \zp{2} \to \zp{1}) = \braket{\zp{2}|\sum_{p} r_p^2|\zp{1}}$ (fm$^2$) of the states in the upper part of Table~\ref{tab:gcm}. The center-of-mass contribution $3/(4A\nu)$ is subtracted from the mean square radii. $M(E0)$ is one half of the isoscalar matrix element under the isospin symmetry. The experimental radius is the point-proton radius derived from the charge radius $2.470\fm$ \cite{Angeli2013}, and $M(E0)$ is from Ref.~\cite{Chernykh2010}. For the Brink-$\alpha$ RXMC, $M(E0) = \sqrt{B(\mathrm{IS}0)}/2$ \cite{Ichikawa2022RXMC}. The radii of Ref.~\cite{Suhara2010} are those of the mass distribution. }
\label{tab:radius}
\small
\begin{tabular}{lcccc}
\toprule
 & $r_{\rm rms}(\zp{1})$ & $r_{\rm rms}(\zp{2})$ & $r_{\rm rms}(\zp{3})$ & $M(E0; \zp{2}\to\zp{1})$ \\
\midrule
\Volkov{}            & 2.33 & 3.33 & 3.92 & 7.4 \\
\Fone{}              & 2.48 & 3.46 & 3.61 & 8.1 \\
\Fthreeb{}           & 2.44 & 3.35 & 3.63 & 6.5 \\
Exp.                 & 2.34 & --- & --- & $5.47 \pm 0.09$ \\
\midrule
\bg-\amd{} (VBP) \cite{Suhara2010}          & 2.31 & 2.90 & 3.26 & --- \\
Brink-$\alpha$ RXMC \cite{Ichikawa2022RXMC} & 2.53 & 4.00 & 4.80 & 10.0 \\
\bottomrule
\end{tabular}
\end{table}

The ground-state radius is $2.33\fm$ for \Volkov, close to the experimental point-proton radius of $2.34\fm$, and $0.11$--$0.15\fm$ larger for \Fone{} and \Fthreeb{} (Table~\ref{tab:radius}). This may be because their three-body terms are repulsive at short distances. The \zp{3} radius is listed for reference, because this state is not converged with respect to the range of $\beta$ of the basis (Sec.~\ref{sec:setup}). 

The \zp{2} radius is larger when the state lies closer to the threshold. It is $3.33$ and $3.35\fm$ for \Volkov{} and \Fthreeb, where the state lies $1.0\MeV$ above the threshold, and $3.46\fm$ for \Fone, where it lies $0.2\MeV$ above it. These radii are $0.9$--$1.0\fm$ larger than the ground-state radii but below the $4.0\fm$ of the Brink $\alpha$-cluster model. The monopole matrix element, $7.4$, $8.1$, and $6.5\fm^2$ for \Volkov, \Fone, and \Fthreeb, exceeds the experimental value $5.47 \pm 0.09\fm^2$ \cite{Chernykh2010} for all three interactions and is closest to it for \Fthreeb.

The rms radius of the \zp{2} state of Ref.~\cite{Suhara2010}, $2.90\fm$, is smaller than the present $3.33\fm$ for \Volkov{} mainly because of the Majorana parameter. With the same Hamiltonian as Ref.~\cite{Suhara2010} ($M = 0.60$), the basis generated with this $M$ gives $2.84\fm$ for the \zp{2} state and $2.23\fm$ for the ground state, against $2.90$ and $2.31\fm$.  The present radii of the \zp{2} state, $3.3$--$3.5\fm$, lie at the lower end of the range of the calculations compiled in Ref.~\cite{Funaki2016}, $3.3$--$4.2\fm$ from the \amd{} \cite{KanadaEnyo2007} and the fermionic molecular dynamics (FMD) \cite{Chernykh2007} to the Tohsaki--Horiuchi--Schuck--R\"opke (THSR) wave function \cite{Funaki2015THSR} and the orthogonality condition model (OCM) \cite{Kurokawa2005}. The value $M(E0) = 6.5\fm^2$ for \Fthreeb{} is consistent with the THSR and FMD values, $6.3$--$6.5\fm^2$, all of which exceed the experimental value.

Thus, once the ground state is fixed, the Hoyle state lies within $0.6\MeV$ of experiment for all three interactions, $1.0\MeV$ above the threshold for \Volkov{} and \Fthreeb{} and $0.2\MeV$ above it for \Fone, against $0.38\MeV$. Its radius, $3.3$--$3.5\fm$, follows the position, and its monopole matrix element, $6.5$--$8.1\fm^2$, exceeds the experimental value for all three interactions. With the same Hamiltonian as Ref.~\cite{Suhara2010} ($M = 0.60$), the variation after the projection gives the ground and \zp{2} states $1.8$ and $1.4\MeV$ lower than the VBP.

\subsection{Correlation excluded by the time-reversal restriction}
\label{sec:trexcluded}
The remaining question about the model space is what the time-reversal restriction of Sec.~\ref{sec:timereversal} excludes and how much it raises the energies of the $0^+$ states. We compare the restricted and unrestricted states at the grid points of the minima of the unrestricted energy surfaces, $(\beta, \gamma) = (0.25, 36^\circ)$ for \Volkov{} and $(0.35, 34^\circ)$ for \Fone, and at $(0.35, 34^\circ)$ also for \Fthreeb. In this comparison, the Majorana parameters are $M = 0.617$ for \Volkov{} and the original values for \Fone{} and \Fthreeb, which differ from those of Table~\ref{tab:force} by at most $0.0069$, and we vary the unrestricted states of the \gcm{} comparison with projection grids coarser than $34^3$.

The two states have nearly the same $0^+$ component, with squared overlaps of $0.97$, $0.98$, and $0.93$ for \Volkov, \Fone, and \Fthreeb, and the unrestricted states have a spin of at most $0.17\hbar$ in each isospin block. They differ in the positions of the paired packets. When we pair the packets of each isospin block so that each pair is closest to a time-reversed pair, the two packets of a pair are $0.34$, $0.61$, and $1.01\fm$ apart in root mean square in the unrestricted states, whereas they coincide in the restricted states. This displacement lowers the energy by $0.82$, $0.24$, and $1.51\MeV$. It lowers the kinetic and Coulomb energies by $2.11$, $3.83$, and $0.78\MeV$ and the three-body energy by $1.41$ (\Fone) and $1.56\MeV$ (\Fthreeb) but raises the central energy by $0.40$, $4.89$, and $2.29\MeV$, mostly in the even channels. The spin-orbit energy changes by $+0.88$, $+0.11$, and $-1.45\MeV$. 

In the \gcm{} with the states of $\beta \le 3.75$, the restriction raises the ground state by $1.14$, $0.79$, and $1.94\MeV$ and the \zp{2} state by $0.22$, $0.09$, and $0.96\MeV$ at the same Majorana parameter. The unrestricted basis has $196$, $193$, and $203$ states instead of $245$, and these increases are therefore lower bounds. Lowering the cutoff to $10^{-5}$ changes these increases by at most $0.05\MeV$. 

To summarize, the restriction excludes a displacement of $0.3$--$1.0\fm$ between the paired packets, which leaves the $0^+$ component nearly unchanged, and raises the ground state by $0.8$--$1.9\MeV$ and the \zp{2} state by $0.09$--$0.96\MeV$, by different amounts for the three interactions. The Majorana parameters of Table~\ref{tab:force} are adjusted in the restricted space, and the comparisons in Sec.~\ref{sec:gcm} and in the next subsection are made within this space. We do not add the unrestricted states to the basis. The basis consists of one state at each grid point whose $0^+$ component is that of the constrained deformation, whereas without the restriction the $0^+$ components at some grid points are close to the $0^+$ state of another deformation (Sec.~\ref{sec:timereversal}). The values above are the energies of a correlation outside the present model space, and whether this correlation is physical remains open.

\subsection{The role of the three-body term}
\label{sec:threebody-role}
We here examine how the three-body term, which provides the saturation property absent from \Volkov, contributes to the energies of the ground state and the Hoyle state. We also examine whether it is the reason that the Hoyle state of \Fone{} lies $0.7\MeV$ below that of \Fthreeb{} (Sec.~\ref{sec:gcm}). Because the Hamiltonian kernel is linear in the strength $V^{(3)}$ of the three-body term, we recompute the kernels without the term on the same basis and evaluate $\braket{\hat V^{(3)}} = V^{(3)}\,dE/dV^{(3)}$ by the Hellmann--Feynman theorem, with the adjusted Majorana parameters. 

Without the three-body term, \Cxii{} collapses, and its ground state lies at least $81.1\MeV$ (\Fthreeb) and $47.0\MeV$ (\Fone) below the \threea{} threshold of the same interaction. In the ground state, the expectation value $\braket{\hat V^{(3)}}$ is $67.9$ and $35.1\MeV$. The three-body term is therefore not a correction to a bound system. It prevents the collapse.

Only the part of $\braket{\hat V^{(3)}}$ in excess of that of three free $\alpha$ particles contributes to the energies measured from the threshold. The value for three free $\alpha$ particles is $31.6$ (\Fthreeb) and $11.7\MeV$ (\Fone). One third to two thirds of the three-body energy arises inside the $\alpha$ particles: $\braket{\hat V^{(3)}}$ is $2.1$ (\Fthreeb) and $3.0$ (\Fone) times that of three free $\alpha$ particles in the ground state, and $1.5$ and $1.8$ times in the \zp{2} state, in which the $\alpha$ clusters still overlap. The excess is $36.3$ and $16.5\MeV$ in the \zp{1} and \zp{2} states for \Fthreeb, and $23.4$ and $9.5\MeV$ for \Fone. The three-body term raises the compact ground state more than the \zp{2} state, and its contribution to the excitation energy is $-19.7\MeV$ for \Fthreeb{} and $-13.8\MeV$ for \Fone.

The rest of the Hamiltonian, i.e., the kinetic energy and the two-body terms, more than compensates this contribution: it contributes $28.0$ (\Fthreeb) and $21.3\MeV$ (\Fone) to the excitation energy. Between the two three-body interactions, the three-body term of \Fthreeb{} lowers the excitation energy by $5.9\MeV$ more than that of \Fone, whereas the Hoyle state of \Fone{} lies $0.73\MeV$ lower. The three-body contribution therefore does not explain the lower Hoyle state of \Fone, which arises in the contribution of the rest of its Hamiltonian, $6.6\MeV$ smaller than that of \Fthreeb.

Thus, the three-body term prevents the collapse of \Cxii{} to at least $47$--$81\MeV$ below the threshold. It raises the compact ground state more than the dilute Hoyle state, and the rest of the Hamiltonian more than compensates its contribution of $-14$ to $-20\MeV$ to the excitation energy. Although the three-body term contributes tens of MeV to the energies, the Hoyle state lies within $0.6\MeV$ of experiment. The lower position for \Fone, $+0.22\MeV$, arises in the contribution of the rest of its Hamiltonian.
\section{Summary}
\label{sec:summary}
In summary, we have proposed a method to evaluate the matrix elements of finite-range three-body interactions between Slater determinants of Gaussian wave packets. With the graph Laplacian of the ranges, as in spectral clustering, the exponential factor of a Gaussian three-body kernel factorizes exactly into pairwise tables, which reduces the number of exponential evaluations from $\mathcal O(A^6)$ to $\mathcal O(A^2)$. A cofactor expansion along one ket index rewrites the remaining contraction as complex matrix--matrix products, and the same factorization applies to the analytic gradient. For \Cxii, the factorization reduces the time of the projected kernels with the three-body term of the Tohsaki F1 interaction from $530$ to $24$ times that of the two-body kernels and makes the \bg-constrained \gcm{} calculation with three-body interactions feasible. Its speed-up for the kernel and its gradient at one Euler node grows from $18$ for \Cxii{} to $93$ for \Caxl.

The variation after the $0^+$ projection under a constraint on the density has a problem that does not depend on the interaction: an intrinsic state with a current or a spin can satisfy the constraint while its small $0^+$ component is close to the $0^+$ state at the energy minimum. We have removed such states by restricting the intrinsic states to the time-reversal-invariant ones. The restriction excludes a displacement of $0.3$--$1.0\fm$ between the two packets of a time-reversed pair and raises the ground state by $0.8$--$1.9\MeV$ and the \zp{2} state by $0.09$--$0.96\MeV$. Whether the excluded correlation is physical remains open. Even with the restriction, the variation after the projection gives the ground and \zp{2} states $1.8$ and $1.4\MeV$ lower than the variation before it with the same Hamiltonian (\Volkov{} with $M = 0.60$).

We have applied the method to the $0^+$ states of \Cxii{} with the \Volkov{}, Tohsaki F1, and Kanada-En'yo--Akaishi F3B interactions, adjusting the Majorana parameter of each interaction to the position of the ground state below the \threea{} threshold. We found that the \zp{2} state lies $0.97$, $0.96$, and $0.22\MeV$ above the threshold for \Volkov, \Fthreeb, and \Fone, respectively, within $0.6\MeV$ of the experimental value of $0.38\MeV$. Its rms radius, $3.3$--$3.5\fm$, follows this position, and its monopole matrix element to the ground state, $6.5$--$8.1\fm^2$, exceeds the experimental value for all three interactions.

The three-body term is not a correction. Without it, the ground state lies at least $47$--$81\MeV$ below the \threea{} threshold of the same interaction. With it, the ground state lies within $1.2\MeV$ of its experimental position below the threshold even with the original Majorana parameters. The three-body term raises the compact ground state more than the dilute \zp{2} state, and the rest of the Hamiltonian more than compensates its contribution to the excitation energy.

We now discuss some points that are not examined in the present calculation. A finite-range three-body interaction is a genuine operator, which can be projected and superposed without an additional prescription. We consider that the fractional power $\rho^{\alpha}$ of energy density functionals is instead an effective mean-field quantity that includes correlations beyond a single Slater determinant, part of which the angular-momentum projection restores explicitly. A three-body interaction calibrated at the projected level therefore need not reproduce the fractional density dependence, and the two approaches should be compared through observables. The three-body terms of the present interactions were not calibrated at this level. Reference~\cite{Tohsaki1994} adjusted F1 to nuclear matter and the $\alpha$-$\alpha$ scattering phase shifts, and Ref.~\cite{KanadaEnyo2004F3B} adjusted the strengths of F3B to closed-shell harmonic-oscillator wave functions and nuclear matter and only its Majorana parameter to the binding energies of the $N = Z$ nuclei from \Heiv{} to \Caxl{} in the projected \amd{} calculation combined with the \gcm. Because \Cxii{} alone cannot constrain the parameters, a consistent determination requires fitting all the parameters to the projected energies of the closed-shell nuclei and $\alpha$-$\alpha$ scattering, which the present method makes computationally feasible.

The factorization applies to Gaussian three-body kernels with any graph of the ranges and to their exchange terms, and extends to $n$-body Gaussian kernels with $n \times n$ determinants of the transition density. Its immediate use is configuration sampling. The quadrupole constraint fixes two collective coordinates, freezes the momenta and spins of the packets at each $(\beta, \gamma)$, and represents the monopole degree of freedom of the \threea{} configurations only through the coarse grid in $\beta$. Sampling the \amd{} configurations without the constraint would decide whether the position of the \zp{2} state above the threshold and its compactness originate from this model space or from the interaction. The RXMC method requires far more kernel evaluations than the \gcm, and the present method provides them for three-body interactions. Apart from the radii and the monopole matrix element, we have not analyzed the wave functions and the densities of the individual $0^+$ states. This analysis with the configurations sampled by the RXMC method will be reported elsewhere. We hope that the present method is helpful to describe cluster states with three-body interactions beyond the constrained model space.

\section*{Acknowledgment}
The author thanks Tadahiro Suhara and Naoyuki Itagaki for valuable comments. The numerical calculations were performed on the heian cluster at YITP, Kyoto University. The author used Claude Code (Anthropic) in writing the analysis code and the manuscript, and takes full responsibility for the content of this paper.

\ifptep\bibliographystyle{ptephy}\else\bibliographystyle{unsrtnat}\fi

\appendix
\section{The inverse overlap matrix as the mean-field density matrix}
\label{app:density}
We show that $\rho = \Bmat^{-1}$ of Sec.~\ref{sec:formalism} is the one-body transition density matrix of the pair $(\Phi, \Phi')$. For $\Phi' = \Phi$, it reduces to the Hartree--Fock density matrix of $\Phi$ \cite{Lowdin1955}.

Let $\Bmat_{ij} = \braket{\psi_i|\psi'_j}$ be nonsingular. We define $\tilde\psi'_i = \sum_l \psi'_l\,\rho_{li}$. These states satisfy $\braket{\psi_j|\tilde\psi'_i} = (\Bmat\rho)_{ji} = \delta_{ji}$. The determinant $\tilde\Phi'$ constructed from the $\tilde\psi'_i$ equals $\Phi'/\det\Bmat$. Therefore, $\braket{\Phi|\hat O|\Phi'}/\braket{\Phi|\Phi'} = \braket{\Phi|\hat O|\tilde\Phi'}$ for any operator. For a one-body operator, the Laplace expansion of the two determinants gives $\sum_{ik}\braket{\psi_i|\hat o|\tilde\psi'_k}\,C_{ik}$. Here $C_{ik}$ are the cofactors of the overlap matrix $\braket{\psi_j|\tilde\psi'_l} = \delta_{jl}$ and hence satisfy $C_{ik} = \delta_{ik}$. We then obtain
\begin{align}
  \frac{\braket{\Phi|\hat O|\Phi'}}{\braket{\Phi|\Phi'}} = \sum_{il}\braket{\psi_i|\hat o|\psi'_l}\,\rho_{li} = \operatorname{tr}(\hat o\,\hat\rho),
  \qquad
  \hat\rho = \sum_{il}\ket{\psi'_l}\,\rho_{li}\,\bra{\psi_i}.
  \label{eq:rhohat}
\end{align}
This is Eq.~\eqref{eq:wick1}. For two- and three-body operators, expanding the determinants similarly yields cofactors given by the $2\times 2$ and $3\times 3$ determinants of Kronecker deltas, leading to Eqs.~\eqref{eq:wick2} and \eqref{eq:wick3}. For any pair, the operator $\hat\rho$ satisfies $\hat\rho^2 = \sum_{lj}\ket{\psi'_l}(\rho\Bmat\rho)_{lj}\bra{\psi_j} = \hat\rho$ and $\operatorname{tr}\hat\rho = \operatorname{tr}(\rho\Bmat) = A$. Thus, it is a projector of rank $A$ onto the space of the ket orbitals along the orthogonal complement of the bra orbitals.

For $\Phi' = \Phi$, $\Bmat$ is Hermitian and positive definite. The L\"owdin orbitals $\varphi_a = \sum_i\psi_i(\Bmat^{-1/2})_{ia}$ are orthonormal and span the occupied space. Hence,
\begin{align}
  \hat\rho = \sum_{ij}\ket{\psi_i}\,(\Bmat^{-1})_{ij}\,\bra{\psi_j} = \sum_{a=1}^{A}\ket{\varphi_a}\bra{\varphi_a}
  \label{eq:rho-hf}
\end{align}
is the Hartree--Fock density matrix of $\Phi$. Its diagonal $\rho(\bm r) = \sum_{ij}\psi_j(\bm r)\rho_{ji}\psi_i^*(\bm r)$ is the mean-field density. Equations~\eqref{eq:wick2} and \eqref{eq:wick3} with $\Phi' = \Phi$ give the Hartree--Fock energies of the two- and three-body interactions, including the exchange terms exactly. For $\Phi' \ne \Phi$, as in the projection and the \gcm, $\hat\rho$ is not Hermitian. Nevertheless, Eqs.~\eqref{eq:wick1}--\eqref{eq:wick3} hold with the same $\rho = \Bmat^{-1}$. The three-body interaction can therefore be projected and superposed. A fractional power of the local density has no corresponding transition form and requires an additional prescription between two determinants.

\section{Derivation of the three-body kernel}
\label{app:kernel3}
We derive Eqs.~\eqref{eq:kappa3}--\eqref{eq:epf} from the wave packet of Eq.~\eqref{eq:packet} in three steps. We first write the product of a bra packet and a ket packet as a Gaussian centered at a complex point. We then write the spatial integral over the three nucleons as a Gaussian integral whose quadratic form is given by the matrix $\mathsf A$ and perform it by completing the square. Finally, we use the unit row sums of $\mathsf G$ to write the remaining exponent in terms of the differences of the transition centers and then in terms of the six packet parameters. The dot product of two vectors does not involve complex conjugation, and $|\bm Z|^2 = \overline{\bm Z}\cdot\bm Z$.

\paragraph{Transition density}
The exponent of Eq.~\eqref{eq:packet} is $-\nu\bm r^2 + 2\sqrt\nu\,\bm Z\cdot\bm r - \bm Z^2/2 - |\bm Z|^2/2$. We add the exponents of the bra packet $\overline{\phi_{\Zi{i}}}$ and the ket packet $\phi_{\bm Z'_l}$ and complete the square in $\bm r$. Since $2\Rcen{il}^2 - (\overline{\Zi{i}}^{\,2} + \bm Z'^{\,2}_l)/2 = \overline{\Zi{i}}\cdot\bm Z'_l$, we obtain
\begin{align}
  \overline{\phi_{\Zi{i}}(\bm r)}\,\phi_{\bm Z'_l}(\bm r)
  = \Big(\frac{2\nu}{\pi}\Big)^{3/2}
  \exp\!\Big[-2\nu\Big(\bm r - \frac{\Rcen{il}}{\sqrt\nu}\Big)^2 + \Wov{il}\Big].
  \label{eq:transition}
\end{align}
The product is a Gaussian with the width parameter $2\nu$ centered at the complex point $\Rcen{il}/\sqrt\nu$. Its integral is the overlap $e^{\Wov{il}}$ of Eq.~\eqref{eq:bmatrix}.

\paragraph{Integration}
The interaction of Eq.~\eqref{eq:general3} does not depend on the spin and isospin. The kernel is therefore the product of the spinor overlaps, the isospin factors, and a spatial integral. We write $\bm s_a = \sqrt\nu\,\bm r_a$ and $\bm R_a = \Rcen{p_a q_a}$, which differs from the packet position $\bm R_i$ of Eq.~\eqref{eq:packet}. By Eq.~\eqref{eq:transition} and $d^3r_a = \nu^{-3/2}d^3s_a$, the spatial integral is
\begin{align}
  V\Big(\frac{2}{\pi}\Big)^{9/2} e^{\Wov{il}+\Wov{jm}+\Wov{kn}}
  \int d^3s_1\,d^3s_2\,d^3s_3\,
  \exp\!\Big[-2\sum_a(\bm s_a - \bm R_a)^2 - \frac{1}{\nu}\sum_{a<b}\kap_{ab}(\bm s_a - \bm s_b)^2\Big].
  \label{eq:integral9}
\end{align}
The second term of the exponent in Eq.~\eqref{eq:integral9} is a sum over the three pairs. Expanding each square, $(\bm s_a - \bm s_b)^2 = \bm s_a^2 + \bm s_b^2 - 2\,\bm s_a\cdot\bm s_b$, we obtain
\begin{align}
  \sum_{a<b}\kap_{ab}(\bm s_a - \bm s_b)^2
  &= (\kap_{12} + \kap_{13})\,\bm s_1^2 + (\kap_{12} + \kap_{23})\,\bm s_2^2 + (\kap_{13} + \kap_{23})\,\bm s_3^2
  \notag\\
  &\quad - 2\kap_{12}\,\bm s_1\cdot\bm s_2 - 2\kap_{13}\,\bm s_1\cdot\bm s_3 - 2\kap_{23}\,\bm s_2\cdot\bm s_3 .
  \notag
\end{align}
The coefficient of $\bm s_a^2$ is the sum of $\kap_{ac}$ over the two pairs that contain $a$, because each pair contributes to the squares of both of its nucleons. We collect the coefficients into a symmetric matrix $\mathsf K$ and write the sum as
\begin{align}
  \sum_{a<b}\kap_{ab}(\bm s_a - \bm s_b)^2 = \sum_{ab}\mathsf K_{ab}\,\bm s_a\cdot\bm s_b .
  \label{eq:laplacian}
\end{align}
The diagonal element $\mathsf K_{aa} = \sum_c\kap_{ac}$ is the coefficient of $\bm s_a^2$. The double sum on the right-hand side contains $\bm s_a\cdot\bm s_b$ twice, as $(a, b)$ and $(b, a)$. The off-diagonal element is therefore one half of the coefficient $-2\kap_{ab}$, which gives $\mathsf K_{ab} = -\kap_{ab}$. This is the graph Laplacian introduced in Sec.~\ref{sec:factorization}, and Eq.~\eqref{eq:laplacian} is its basic identity \cite{Luxburg2007}. Setting $\kap_{12} = \kap_{13} = \kap$, and $\kap_{23} = \kap$ for the triangle topology or $\kap_{23} = 0$ for the apex topology with the first slot as the apex, we obtain
\begin{align}
  \mathsf K_{\rm triangle} = \kap\begin{pmatrix} 2&-1&-1\\ -1&2&-1\\ -1&-1&2 \end{pmatrix},
  \qquad
  \mathsf K_{\rm apex} = \kap\begin{pmatrix} 2&-1&-1\\ -1&1&0\\ -1&0&1 \end{pmatrix}.
  \label{eq:K-matrices}
\end{align}
Each row of $\mathsf K$ sums to zero, $\mathsf K\bm 1 = 0$ for $\bm 1 = (1, 1, 1)$, because a common translation of the three nucleons does not change their relative coordinates. The matrix $\mathsf K$ is positive semidefinite, because the left-hand side of Eq.~\eqref{eq:laplacian} is not negative. For a $3\times 3$ matrix $\mathsf M$ acting on the particle index, we write $\bm s^{\mathsf T}\mathsf M\bm s' = \sum_{ab}\mathsf M_{ab}\,\bm s_a\cdot\bm s'_b$. With Eq.~\eqref{eq:laplacian}, the exponent of Eq.~\eqref{eq:integral9} is $-2[\bm s^{\mathsf T}\mathsf A\,\bm s - 2\bm R^{\mathsf T}\bm s + \bm R^{\mathsf T}\bm R]$ with $\mathsf A = 1 + \mathsf K/2\nu$ of Eq.~\eqref{eq:kappa3}. The matrix $\mathsf A$ is symmetric and positive definite, and its eigenvalues are not smaller than unity. Because $\mathsf G = \mathsf A^{-1}$ is also symmetric and $\mathsf G\mathsf A = 1$, expanding the right-hand side shows
\begin{align}
  \bm s^{\mathsf T}\mathsf A\,\bm s - 2\bm R^{\mathsf T}\bm s
  = (\bm s - \mathsf G\bm R)^{\mathsf T}\mathsf A\,(\bm s - \mathsf G\bm R) - \bm R^{\mathsf T}\mathsf G\bm R ,
  \notag
\end{align}
and the exponent becomes
\begin{align}
  -2(\bm s - \mathsf G\bm R)^{\mathsf T}\mathsf A\,(\bm s - \mathsf G\bm R)
  - 2\bm R^{\mathsf T}(1 - \mathsf G)\,\bm R .
  \label{eq:completed}
\end{align}
The first term is a Gaussian in $\bm s$ centered at $\mathsf G\bm R$, and the second term does not depend on $\bm s$.
For a real symmetric positive-definite $n\times n$ matrix $\mathsf M$ and a complex vector $\bm c$,
\begin{align}
  \int_{\mathbb R^n} d^n s\, e^{-(\bm s - \bm c)^{\mathsf T}\mathsf M(\bm s - \bm c)} = \frac{\pi^{n/2}}{\sqrt{\det\mathsf M}} .
  \label{eq:gauss}
\end{align}
A linear transformation of variables that reduces $\mathsf M$ to the identity matrix gives the factor $(\det\mathsf M)^{-1/2}$ and a product of $n$ one-dimensional integrals $\int_{-\infty}^{\infty}dt\,e^{-(t - d)^2} = \sqrt\pi$. This holds also for complex $d$, as follows from Eqs.~7.4.32 and 7.1.16 of Ref.~\cite{Abramowitz1964}. We apply Eq.~\eqref{eq:gauss} to the first term of Eq.~\eqref{eq:completed} for each Cartesian component, with $n = 3$ and $\mathsf M = 2\mathsf A$. The three components give $(\pi/2)^{9/2}(\det\mathsf A)^{-3/2}$, and the prefactor becomes $\widetilde V = V(\det\mathsf A)^{-3/2}$.

\paragraph{Exponent}
We now rewrite the second term of Eq.~\eqref{eq:completed}. Since $\mathsf K\bm 1 = 0$, we have $\mathsf A\bm 1 = \bm 1$ and hence $\mathsf G\bm 1 = \bm 1$, so that all row sums of $\mathsf G$ equal unity. The matrix $1 - \mathsf G$ is thus symmetric with zero row sums. A symmetric matrix $\mathsf M$ with zero row sums satisfies $\mathsf M_{aa} = -\sum_{b\ne a}\mathsf M_{ab}$ and therefore has the form of $\mathsf K$ with $\kap_{ab} = -\mathsf M_{ab}$ for $a \ne b$. Equation~\eqref{eq:laplacian} then gives
\begin{align}
  \sum_{ab}\mathsf M_{ab}\,\bm R_a\cdot\bm R_b = -\sum_{a<b}\mathsf M_{ab}\,(\bm R_a - \bm R_b)^2 .
  \label{eq:rowsum}
\end{align}
We apply Eq.~\eqref{eq:rowsum} to $\mathsf M = 1 - \mathsf G$. The right-hand side contains only the off-diagonal elements, which are $\mathsf M_{ab} = -\mathsf G_{ab}$. The second term of Eq.~\eqref{eq:completed} therefore becomes
\begin{align}
  -2\bm R^{\mathsf T}(1 - \mathsf G)\,\bm R = -2\sum_{a<b}\mathsf G_{ab}\,(\bm R_a - \bm R_b)^2 .
  \notag
\end{align}
Together with the overlaps and the spin and isospin factors, this gives Eq.~\eqref{eq:kernel3}. The exponent depends only on the differences of the transition centers, because $\hat v$ depends only on the relative coordinates.

\paragraph{The matrix $\mathsf G$}
The kernel of Eq.~\eqref{eq:kernel3} requires $\mathsf G$ in the exponent and $\det\mathsf A$ in $\widetilde V$. We obtain both from the eigenvalues of $\mathsf K$. Let $\lambda_\mu$ and $\bm u_\mu$ be the eigenvalues and the normalized eigenvectors of $\mathsf K$. The matrix $\mathsf A = 1 + \mathsf K/2\nu$ has the same eigenvectors with the eigenvalues $1 + \lambda_\mu/2\nu$, and its inverse $\mathsf G$ has the eigenvalues $2\nu/(2\nu + \lambda_\mu)$. Therefore,
\begin{align}
  \mathsf G = \sum_{\mu}\frac{2\nu}{2\nu + \lambda_\mu}\,\bm u_\mu\bm u_\mu^{\mathsf T},
  \qquad
  \det\mathsf A = \prod_{\mu}\Big(1 + \frac{\lambda_\mu}{2\nu}\Big).
  \label{eq:G-spectral}
\end{align}
The eigenvectors are the Jacobi coordinates of the three nucleons. The center-of-mass vector $\bm 1/\sqrt3$ has $\lambda = 0$ for both topologies, and its factor $2\nu/(2\nu + \lambda)$ is unity. This is the unit row sum $\mathsf G\bm 1 = \bm 1$ used above. In the Jacobi coordinates, $\mathsf G$ is diagonal. It leaves the center of mass of the three transition centers unchanged and reduces each relative coordinate by the factor $2\nu/(2\nu + \lambda_\mu)$. The center $\mathsf G\bm R$ of the Gaussian in Eq.~\eqref{eq:completed} is thus closer to the center of mass than $\bm R$, because the Gaussian of the interaction weights the configurations in which the three nucleons are close to each other. For the triangle topology, the other two eigenvalues are $3\kap$, and
\begin{align}
  \mathsf G_{\rm triangle} = \frac{1}{2\nu + 3\kap}
  \begin{pmatrix} 2\nu+\kap & \kap & \kap\\ \kap & 2\nu+\kap & \kap\\ \kap & \kap & 2\nu+\kap \end{pmatrix}.
  \notag
\end{align}
For the apex topology with the first slot as the apex, $(0, 1, -1)/\sqrt2$ has $\lambda = \kap$ and $(2, -1, -1)/\sqrt6$ has $\lambda = 3\kap$. Equation~\eqref{eq:G-spectral} gives $2\mathsf G_{12} = 2\mathsf G_{13} = 2\kap/(2\nu + 3\kap)$ and $2\mathsf G_{23} = 2\kap^2/[(2\nu + \kap)(2\nu + 3\kap)]$, and the diagonal elements follow from the unit row sums. Although $\kap_{23} = 0$, $\mathsf G_{23}$ does not vanish. The two nucleons attached to the apex are correlated through it, and this correlation is of second order in $\kap$. For $\kap \ll \nu$, the weights $2\nu\mathsf G_{12} = 2\nu\mathsf G_{13}$ reduce to $\kap$, as in the triangle topology. For the apex topology, $\det\mathsf A = (2\nu + \kap)(2\nu + 3\kap)/(2\nu)^2$.

\paragraph{Pairwise form}
By Eq.~\eqref{eq:rowsum}, the exponent of Eq.~\eqref{eq:kernel3} is $-2\sum_{ab}(\delta_{ab} - \mathsf G_{ab})\,\bm R_a\cdot\bm R_b + \sum_a\Wov{p_a q_a}$. We substitute $\bm R_a = (\overline{\bm Z}_{p_a} + \bm Z'_{q_a})/2$ into the first sum. Because $\mathsf G$ is symmetric, the two bra-ket terms of $\bm R_a\cdot\bm R_b$ give the same sum, and
\begin{align}
  -2\sum_{ab}(\delta_{ab} - \mathsf G_{ab})\,\bm R_a\cdot\bm R_b
  &= -\frac12\sum_{ab}(\delta_{ab} - \mathsf G_{ab})\big(\overline{\bm Z}_{p_a}\!\cdot\overline{\bm Z}_{p_b} + \bm Z'_{q_a}\!\cdot\bm Z'_{q_b}\big)
  \notag\\
  &\quad - \sum_{ab}(\delta_{ab} - \mathsf G_{ab})\,\overline{\bm Z}_{p_a}\!\cdot\bm Z'_{q_b} .
  \notag
\end{align}
The overlaps give $\sum_a\Wov{p_a q_a} = \sum_a\overline{\bm Z}_{p_a}\!\cdot\bm Z'_{q_a} - \frac12\sum_a\big(|\bm Z_{p_a}|^2 + |\bm Z'_{q_a}|^2\big)$. The part with $\delta_{ab}$ in the last sum cancels the first term of the overlaps, and the remaining terms give Eq.~\eqref{eq:epf}.

\paragraph{Exchange terms of the apex topology}
Expanding the products in Eq.~\eqref{eq:tohsaki} gives five spatial permutations: the identity, $\hat P^r_{ij}$, $\hat P^r_{ik}$, $\hat P^r_{ij}\hat P^r_{ik}$, and $\hat P^r_{ik}\hat P^r_{ij}$, with the coefficients $(W^{(3)}_\alpha)^2$, $W^{(3)}_\alpha M^{(3)}_\alpha$, $W^{(3)}_\alpha M^{(3)}_\alpha$, $(M^{(3)}_\alpha)^2/2$, and $(M^{(3)}_\alpha)^2/2$. Because $\hat P^r$ exchanges the spatial coordinates only, a permutation $\pi$ replaces $q_a$ by $q_{\pi(a)}$ in $\Rcen{p_a q_a}$ and $\Wov{p_a q_a}$ of Eq.~\eqref{eq:kernel3}. The spinor overlaps and the isospin factors remain those of the pairs $(p_a, q_a)$. For example, $\hat P^r_{ij}$ exchanges the spatial coordinates of the first two nucleons, so that the spatial factors are evaluated with the pairs $(p_1, q_2)$ and $(p_2, q_1)$, whereas the spinor overlaps remain those of $(p_1, q_1)$ and $(p_2, q_2)$. Each term thus has the form of Eq.~\eqref{eq:epf} with permuted ket indices and uses the same tables. The other choices of the apex are relabelings of the slots. They give the same contribution to Eq.~\eqref{eq:wick3}, because $\mathcal D_{ijk,lmn}$ is invariant under a common permutation of the bra and ket slots.

\section{Gradient of the three-body term}
\label{app:gradient}
The variation of Sec.~\ref{sec:intrinsic} requires the gradient of the projected energy. We treat $\Zi{i}$ and $\overline{\Zi{i}}$, as well as the spinor amplitudes and their complex conjugates, as independent variables. Here we give the derivative of the three-body term with respect to $\overline{\Zi{h}}$. The determinant $\mathcal D_{ijk,lmn}$ of Eq.~\eqref{eq:wick3} is multilinear in its rows, and $\partial\rho = -\rho\,(\partial\Bmat)\,\rho$. Therefore,
\begin{align}
  \partial \frac{\braket{\Phi|\hat V^{(3)}|\Phi'}}{\braket{\Phi|\Phi'}}
  = \frac{1}{3!}\sum_{ijk}\sum_{lmn}\Big[(\partial V_{ijk,lmn})\,\mathcal D_{ijk,lmn}
  + V_{ijk,lmn}\sum_{t=1}^{3}\mathcal D^{(t)}_{ijk,lmn}\Big].
  \label{eq:grad3}
\end{align}
Here $\mathcal D^{(t)}$ denotes $\mathcal D$ with its $t$-th row replaced by the corresponding row of $\partial\rho$. The prefactor $\widetilde V$ and the matrix $\mathsf G$ of Eq.~\eqref{eq:kappa3} do not depend on the packets. The derivative of the kernel in Eq.~\eqref{eq:kernel3} is therefore the kernel multiplied by the derivative of its exponent,
\begin{align}
  \frac{\partial V_{ijk,lmn}}{\partial\overline{Z}^{\,\alpha}_h}
  &= V_{ijk,lmn}\Big[-\sum_{a<b}2\mathsf G_{ab}\,\big(\Rcen{p_a q_a} - \Rcen{p_b q_b}\big)^{\alpha}\big(\delta_{p_a h} - \delta_{p_b h}\big)
  + \sum_{a=1}^{3}\frac{\partial\Wov{p_a q_a}}{\partial\overline{Z}^{\,\alpha}_h}\Big],
  \notag\\
  \frac{\partial\Wov{rs}}{\partial\overline{Z}^{\,\alpha}_h}
  &= \delta_{rh}\Big(Z'^{\,\alpha}_s - \frac{Z^{\alpha}_r}{2}\Big) - \frac{\delta_{sh}}{2}\,Z^{\alpha}_s .
  \label{eq:dkernel3}
\end{align}
Here $\alpha$ denotes a Cartesian component. The last term of $\partial\Wov{rs}$ is the derivative of the normalization of the ket packet. It is present when the ket is the bra itself or its rotated image, as in the projection. This term generates an overall rescaling of the ket packet, under which the energy is invariant. We omit it consistently from both $\partial V$ and $\partial\Bmat$.

For a fixed bra triple and $h \in \{i,j,k\}$, each term in the bracket of Eq.~\eqref{eq:dkernel3} depends on one ket slot. The explicit term in Eq.~\eqref{eq:grad3} is therefore a sum of three contractions of the form of Eq.~\eqref{eq:fixed-bra}. In each contraction, one $X_a(q)$ is multiplied by a linear function of $\bm Z'_q$. The implicit term replaces one row of $\mathcal D$ by the corresponding row of $\partial\rho$. This replacement enters through $Y_{n,c}$ of Eq.~\eqref{eq:gemm3} or the minor $\mathcal C_c$ of Eq.~\eqref{eq:gemm3-sum}. All terms share the left operand $T$ of Eq.~\eqref{eq:gemm3}. The gradient requires a few contractions comparable in size to the energy contraction.

In the variation after the $J^\pi = 0^+$ projection, the norm and Hamiltonian kernels with the rotated ket are $\mathcal N(\Omega) = \det\Bmat(\Omega)$ and $\mathcal H(\Omega) = \det\Bmat(\Omega)\,E(\Omega)$. Jacobi's formula gives $\partial\mathcal N = \mathcal N\,\operatorname{tr}(\rho\,\partial\Bmat)$. We integrate the kernel gradients with the same quadrature as the kernels. The gradient of the projected energy $E^J = H^J/N^J$ is $\partial E^J = (\partial H^J - E^J\,\partial N^J)/N^J$. We have confirmed that it agrees with a finite difference evaluated using the same quadrature for an extended configuration including the three-body term. 
\section{Two-body matrix elements}
\label{app:twobody}
This appendix gives the kernels of the central, spin-orbit, and Coulomb interactions in the notation of Sec.~\ref{sec:factorization}, and the form of their contraction used for the two-body kernels in Sec.~\ref{sec:performance}. We use the pair variable $\bm r_{il,jm} = \Rcen{il} - \Rcen{jm}$ for the bra indices $i, j$ and the ket indices $l, m$. The spatial part of the central kernel of Eq.~\eqref{eq:central} is
\begin{align}
  f^{\rm cent}_{ij,lm} &= \sum_\alpha \widetilde V^{(2)}_\alpha\,
  \exp\!\big[-\widetilde\kap_\alpha\,\bm r_{il,jm}^2 + \Wov{il} + \Wov{jm}\big],
  \label{eq:f-cent}\\
  \widetilde V^{(2)}_\alpha &= V^{(2)}_\alpha\Big(\frac{\nu}{\nu + \kap_\alpha}\Big)^{3/2},
  \qquad
  \widetilde\kap_\alpha = \frac{\kap_\alpha}{\nu + \kap_\alpha}.
  \notag
\end{align}
Equation~\eqref{eq:f-cent} is the two-particle case of Eq.~\eqref{eq:kernel3} (Appendix~\ref{app:kernel3}). For two particles, $\mathsf K = \kap_\alpha\bigl(\begin{smallmatrix} 1&-1\\ -1&1 \end{smallmatrix}\bigr)$, $\det\mathsf A = 1 + \kap_\alpha/\nu$, and $2\mathsf G_{12} = \kap_\alpha/(\nu + \kap_\alpha)$, which give $\widetilde V^{(2)}_\alpha$ and $\widetilde\kap_\alpha$. In terms of the centers $\Rcen{il}/\sqrt\nu$ of the transition densities, the range parameter of the exponent is $\kap_\alpha\nu/(\nu + \kap_\alpha)$, i.e., $1/\kap_\alpha + 1/\nu$ in its inverse. This is the folding of the Gaussian interaction with the distribution $e^{-\nu r^2}$ of the relative coordinate of two packets.

The spin factor of the central kernel has two channels. With the overlap $s_{il} = \braket{\chi_i|\chi'_l}$ of two spinors, the direct channel multiplies Eq.~\eqref{eq:f-cent} by $W_\alpha s_{il}s_{jm} + B_\alpha s_{im}s_{jl}$, and the exchange channel, in which the two ket packets are interchanged, by $M_\alpha s_{il}s_{jm} + H_\alpha s_{im}s_{jl}$. The isospin is diagonal in the packet basis. We therefore evaluate the $nn$, $pp$, and $np$ sectors separately. The exchange terms of Eq.~\eqref{eq:wick2}, generated by the same kernel with $l \leftrightarrow m$, supply the antisymmetrization between identical nucleons.

The spin-orbit kernel has the Gaussian factor of Eq.~\eqref{eq:f-cent} with the ranges $\kap^{\rm LS}_\alpha$, multiplied by the matrix elements of the relative orbital angular momentum and of the total spin,
\begin{align}
  f^{\rm LS}_{ij,lm} &= \sum_\alpha \widetilde u_\alpha\,
  \exp\!\big[-\widetilde\kap^{\rm LS}_\alpha\,\bm r_{il,jm}^2 + \Wov{il} + \Wov{jm}\big]\,
  \bm L_{ij,lm}\cdot\bm S_{ij,lm},
  \notag\\
  \widetilde u_\alpha &= u_\alpha\Big(\frac{\nu}{\nu + \kap^{\rm LS}_\alpha}\Big)^{5/2},
  \qquad
  \bm L_{ij,lm} = -\frac{i}{2}\,\overline{\bm d}_{ij}\times\bm d_{lm},
\end{align}
where $\bm d_{lm} = \bm Z'_l - \bm Z'_m$, $\overline{\bm d}_{ij} = \overline{\Zi{i}} - \overline{\bm Z}_j$, and $\bm S_{ij,lm} = \tfrac12\big(\bm\sigma_{il}\,s_{jm} + s_{il}\,\bm\sigma_{jm}\big)$ is the matrix element of the total spin, with $\bm\sigma_{il} = \braket{\chi_i|\bm\sigma|\chi'_l}$. For identical nucleons, the antisymmetrization of Eq.~\eqref{eq:wick2} restricts the spin-orbit kernel to the triplet-odd channel. For a neutron--proton pair, we include the isospin part of $\hat P({}^3{\rm O})$ explicitly. It gives one half of the difference between the direct channel and the channel in which the spatial and spin parts of the two ket packets are interchanged.

For a proton pair, the spatial part of the Coulomb kernel, which multiplies $s_{il}s_{jm}$, is the Coulomb interaction folded with the same distribution $e^{-\nu r^2}$,
\begin{align}
  f^{\rm Coul}_{ij,lm} = e^2\sqrt\nu\,\frac{\operatorname{erf}\big(\sqrt{\bm r_{il,jm}^2}\big)}{\sqrt{\bm r_{il,jm}^2}}\,e^{\Wov{il} + \Wov{jm}},
\end{align}
where the square root of the complex $\bm r^2_{il,jm}$ is taken on the principal branch and the ratio is analytic at $\bm r^2_{il,jm} = 0$. When the transition centers are far apart compared with $1/\sqrt\nu$, the error function approaches unity, and the kernel reduces to $e^2$ divided by the distance between the centers, multiplied by the overlaps.

The contraction of the Gaussian kernels is written as matrix products in the same way as that of the three-body kernel. The four indices are connected by six pairwise factors, and the contraction of Eq.~\eqref{eq:wick2} remains $\mathcal O(A^4)$ for any order of summation. We multiply a bra--ket table elementwise by the transition density and fix one bra index, so that the sum over the ket indices becomes an $A \times A$ matrix product, and stack these products over the fixed index into a single $(A \times A)(A \times A^2)$ product. The factors that depend only on the bra pair or only on the ket pair, such as $\exp[-\widetilde\kap\,\bm d_{lm}^2/4]$, do not change under the rotation of the ket packets, because $\bm d\cdot\bm d$ is a bilinear form without complex conjugation, and we evaluate them once per pair of basis functions. The Coulomb kernel does not factorize in this way, because its error function is not a product of pairwise factors. In the measurements of Sec.~\ref{sec:performance}, the central kernels are contracted in this form, and the spin-orbit and Coulomb kernels are summed directly.

\end{document}